\documentclass[AMA,STIX1COL]{MyWileyNJD-v2}
\usepackage{moreverb}

\newcommand\BibTeX{{\rmfamily B\kern-.05em \textsc{i\kern-.025em b}\kern-.08em
T\kern-.1667em\lower.7ex\hbox{E}\kern-.125emX}}

\newcommand{\Var}{\mathrm{Var}}

\articletype{A Preprint}%

\begin{document}

\title{Natural cubic splines for the analysis of Alzheimer's clinical
trials}

\author[1]{M.C. Donohue*}

\author[1]{O. Langford}

\author[2]{P. Insel}

\author[3]{C.H. van Dyck}

\author[4]{R.C. Petersen}

\author[5]{S. Craft}

\author[1]{G. Sethuraman}

\author[1]{R. Raman}

\author[1]{P.S. Aisen}

\author{For the Alzheimer's Disease Neuroimaging Initiative\(^{**}\)}

\authormark{DONOHUE \textsc{et al}}

\address[1]{\orgdiv{Alzheimer's Therapeutic Research Institute},
\orgname{University of Southern California}
\orgaddress{\state{California}, \country{USA}}}

\address[2]{\orgdiv{Department of Psychiatry}, 
\orgname{University of California, San Francisco}, 
\orgaddress{\state{California}, \country{USA}}}

\address[3]{\orgdiv{Alzheimer's Disease Research Unit}, 
\orgname{Yale School of Medicine}, 
\orgaddress{\state{Connecticut}, \country{USA}}}

\address[4]{\orgname{Mayo Clinic, Rochester}, 
\orgaddress{\state{Minnesota}, \country{USA}}}

\address[5]{\orgdiv{Department of Internal Medicine--Geriatrics}, 
\orgname{Wake Forest School of Medicine}, 
\orgaddress{\state{North Carolina}, \country{USA}}}

\corres{*Michael C. Donohue, Alzheimer's Therapeutic Research
Institute, University of Southern California, 9860 Mesa Rim Rd, San
Diego, CA 92121, USA. \email{mdonohue@usc.edu}}

\abstract{Mixed model repeated measures (MMRM) is the most common analysis
approach used in clinical trials for Alzheimer's disease and other
progressive diseases measured with continuous outcomes measured over
time. The model treats time as a categorical variable, which allows an
unconstrained estimate of the mean for each study visit in each
randomized group. Categorizing time in this way can be problematic when
assessments occur off-schedule, as including off-schedule visits can
induce bias, and excluding them ignores valuable information and
violates the intention to treat principle. This problem has been
exacerbated by clinical trial visits which have been delayed due to the
COVID19 pandemic. As an alternative to MMRM, we propose a constrained
longitudinal data analysis with natural cubic splines that treats time
as continuous and uses test version effects to model the mean over time.
The spline model is shown to be superior, in practice and simulation
studies, to categorical-time models like MMRM and models that assume a
proportional treatment effect.}


\maketitle

\footnotesize

\(^{**}\)Data used in preparation of this article were obtained from the
Alzheimer's Disease Neuroimaging Initiative (ADNI) database
(adni.loni.usc.edu). As such, the investigators within the ADNI
contributed to the design and implementation of ADNI and/or provided
data but did not participate in analysis or writing of this report. A
complete listing of ADNI investigators can be found at:
\url{http://adni.loni.usc.edu/wp-content/uploads/how_to_apply/ADNI_Acknowledgement_List.pdf}

\normalsize

\providecommand{\tightlist}{%
  \setlength{\itemsep}{0pt}\setlength{\parskip}{0pt}}

\hypertarget{background}{%
\section{Background}\label{background}}

The mixed model repeated measures (MMRM)
\citep{mallinckrodt2001accounting} is a common analytic approach for
randomized clinical trials with continuous outcomes collected
longitudinally at several follow-up visits. The MMRM generally assumes
that the mean change from baseline in the outcome is a linear function
of the baseline value of the outcome, time (as a categorical variable
reflecting timing of study visits), and the time-by-treatment
interaction. This allows an estimate of the mean at each visit for both
groups without assuming any trend over time. The primary test statistic
is typically the estimated difference between group means at the final
visit. The model generally includes no random effects. Instead residuals
are assumed to be correlated from visit-to-visit within participant. The
MMRM is virtually assumption-free with respect to the shape of the mean
over time and the correlation structure, which likely explains its broad
use in pharmaceutical statistics.

However, the MMRM requires many parameters. If \(K\) is the number of
visits, MMRM requires \(2K\) mean parameters, \(K(K-1)/2\) correlation
parameters (assuming a general symmetric correlation structure), and
\(K\) variance parameters (assuming heterogeneous variance). In this
respect, MMRM might be seen to be favoring reduced bias at the cost of
statistical efficiency and power.

Another concern is that clinical trial visits often occur off-schedule.
Study protocols generally allow a visit window (e.g.~visits may occur
within two weeks of a target date). When observations fall outside these
windows, investigators are forced to decide between ignoring data or
including it and introducing potential bias due to observations being
effectively carried forward or backward to the closest target date
\citep{andersen2013practical}. The problem of late visits can be
exacerbated by unforeseen interruptions, such as those caused by the
COVID19 pandemic. In these situations, it might be advantageous to treat
time as continuous, thus allowing the model to be informed by the actual
time from baseline, rather than categorized time.

An alternative to MMRM was proposed for clinical trials in
autosomal‐dominant Alzheimer's disease
\citep{bateman2017dian, wang2018novel}. The time variable is the
expected year of symptom onset (EYO) and mean in the placebo group is
modeled as a monotonically decreasing step function of EYO. The
treatment benefit is assumed to be \emph{proportional} to the decline,
making the model nonlinear. The model was extended to a multiple outcome
model for the final analysis, but the monotonicity and proportional
treatment effect assumptions were violated and the model failed to
converge \citep{salloway2021trial}.

In this paper we explore another alternative model for Alzheimer's
disease trials, one that treats time as continuous. We consider natural
cubic splines \citep{hastie1992natural} to flexibly model the temporal
mean trend with fewer parameters than MMRM. We compare the natural cubic
splines model to categorical-time models (MMRM) and models assuming a
proportional treatment effect in three completed Alzheimer's studies and
in simulations. We also explore the utility of adjusting for alternating
cognitive test versions, which are known to vary in difficulty. A
simulation study is presented to demonstrate the power and Type I error
of each approach. The spline model is shown to provide superior power
with good control of Type I error.

\hypertarget{model-specifications}{%
\section{Model specifications}\label{model-specifications}}

\textbf{Categorical-time mean structure.} The first type of mean
structure treats time as a categorical variable, as in the MMRM. The
response for subject \(i\) and study visit \(j=1,\ldots,K\) is modeled
as:

\begin{align*}
Y_{ij} = \beta_0 + &\beta_1 1\{j=1\} + \beta_2 1\{j=2\} + \cdots + \beta_K 1\{j=K\} +  \\
                  (&\gamma_2 1\{j=2\} + \gamma_3 1\{j=2\} + \cdots + \gamma_K 1\{j=K\})\textrm{Active}_i + \varepsilon_{ij}
\end{align*}

where ``Active\(_i\)'' is 1 if subject \(i\) is in the active group and
0 otherwise. The \(\beta\) terms express the mean at each visit in the
placebo group (e.g.~the estimated mean at the second visit is
\(\hat\beta_0+\hat\beta_2\)) and the \(\gamma\) terms represent the
treatment group difference at each visit. We focus here on the temporal
mean structure, but other baseline covariate effects can be added with
additional terms as usual. Note that the exclusion of a \(\gamma_1\)
term constrains the mean at baseline to be the same for both groups.
This is the constraint referred to in constrained longitudinal data
analysis (cLDA) \citep{liang2000longitudinal}. This categorical-time
mean structure is similar to MMRM, but in MMRM the response variable is
the change from baseline, \(Y^*_{ij}=Y_{ij}-Y_{i1}\), for
\(j=2,\ldots,L\); the baseline score \(Y_{i1}\) is included as a
covariate; and the \(\beta_1 1\{j=1\}\) term is dropped. The cLDA
parameterization is more efficient than MMRM when baseline observations
are subject to missingness \citep{lu2010efficiency}. We explore several
different variance-covariance structures for the residuals
\(\varepsilon_{ij}\) as described at the end of this section.

\textbf{Linear mean structure.} Alternatively, time from baseline at
visit \(j\) can be treated as a continuous variable \(t_j\). The linear
mean structure assumes:

\[
Y_{ij} = \beta_0 + \beta_1t_j + \gamma t_j\textrm{Active}_i + \varepsilon_{ij}.
\]

The \(\beta_0\) term represents the mean in both groups at baseline,
\(\beta_1\) represents the rate of change per month in the control group
and the \(\gamma\) term represents the difference in slopes.

\textbf{Natural cubic spline mean structure.} The natural cubic spline
\citep{hastie1992natural} structure also treats time as continuous. A
cubic spline is a function defined by cubic polynomials that are spliced
together at knot locations and the resulting function is restricted to
be continuous and have continuous first and second derivatives. A
natural cubic spline has a further constraint of having a second
derivative of zero at the boundaries (i.e.~acceleration of the curve is
not allowed at the first and last observation). A spline function
\(f(t)\) can be expressed as a linear combination of basis functions,
\(b_k(t)\), and weights \(\beta_k\):
\(f(t) = \beta_0 + \sum_{k=1}^m\beta_kb_k(t)\). The \(b_k(t)\) basis
functions can be numerically determined given the boundaries, interior
knot locations, and the continuity and derivative constraints. In
practice, this a processing step that does not involve the observed
outcomes, akin to calculating \(t^2\), \(t^3\), etc. for polynomial
regression. As with polynomial regression, once the \(b_k(t)\) are
determined, the \(\beta_k\) weights can be estimated by maximum
likelihood estimation.

In our clinical trial context, the natural cubic spline model assumes:

\[
Y_{ij} = \beta_0 + \sum_{k=1}^m\beta_kb_k(t_j) + \textrm{Active}_i\sum_{k=1}^m\gamma_k b_k(t_j) + \varepsilon_{ij}.
\] where \(b_k(t)\) are the known spline basis functions for a given set
of knots and extreme values of \(t\). Interior knots are typically
spaced according to quantiles of \(t\). The resulting curve for the
placebo group is defined by the natural cubic spline
\(f(t) = \beta_0 + \sum_{k=1}^m\beta_kb_k(t_j)\); while the natural
cubic spline \(g(t)=\sum_{k=1}^m\gamma_k b_k(t_j)\) represents the
treatment group difference over time and is constrained to be zero at
time zero. The number of interior knots (m-2) needs to be specified.
Given the known basis functions, estimation of the unknown parameters
(\(\beta\)s and \(\gamma\)s) is accomplished with generalized least
squares as with the prior two models.

Again, as with the other temporal mean structures, additional covariates
can be added. In particular, we will add a time-varying covariate for
cognitive test version. To avoid test familiarity, the typical
Alzheimer's clinical trial alternates cognitive tests among three
versions with different word lists or stories to be recalled. The
versions are intended to be similar in difficulty, but differences can
be detected. The time-varying covariate effect can be interpreted as a
version difficulty offset and can be added to the models that assume a
suitably smooth parametric tend over time. We also explore adding the
test version effect to categorical-time models, however, we would not
expect this to have a large impact because test versions alternate with
the visit categories by design.

Cognitive test version at time \(t\) is clearly an exogenous variable,
independent of values of the assessment before time \(t\), and should
not bias or otherwise interfere with the interpretation of the treatment
effects \citep{diggle2002analysis}. The schedule of test versions is
typically entirely determined by the protocol and any deviations from
the protocol are due to administration errors unrelated to prior test
performance. Nevertheless, an alternative strategy would be to adjust
the scoring rules for the test version's difficulty level using data
external to the trial.

\textbf{Proportional treatment effect.} The proportional treatment
effect model is of the form

\[
Y_{ij} = \beta_0 + (\beta_1 1\{j=1\} + \beta_2 1\{j=2\} + \cdots + \beta_K 1\{j=K\})\exp(\theta \textrm{Active}_i) + \alpha_i + \varepsilon_{ij}
\]

where \(\alpha_i\) are participant-specific random intercepts assumed to
be distributed \(\mathcal{N}(0,\sigma_\alpha)\), and
\(\varepsilon_{ij}\) are assumed to independent and identically
distributed \(\mathcal{N}(0,\sigma)\). The mean structure assumption for
the placebo group is identical to the categorical-time model above, but
the active group is assumed to differ at every visit by a constant
factor of \(\exp(\theta)\).

\textbf{Variance-covariance assumptions}. Since the mean structures in
this paper treat time as either categorical or continuous, we explored
variance-covariance assumptions under both paradigms as well. The
considered variance-covariance assumptions include:

\begin{itemize}
\tightlist
\item
  \textbf{Unstructured}: The vector of residuals for individual \(i\)
  \(\varepsilon_{i}\) are assumed to follow a multivariate Gaussian
  distribution with mean zero and general symmetric correlation and
  heterogeneous variance per study visit:
  \(\varepsilon_{i} \sim, \mathcal{N}(0, V\Sigma V)\) for diagonal
  matrix \(V\) and symmetric general matrix \(\Sigma\),
\item
  \textbf{``AR1 Het.''}: \emph{categorical-time} autoregressive order
  one correlation with heterogeneous variance per visit,
\item
  \textbf{``CAR1 Const. Prop.''}: \emph{continuous-time} autoregressive
  order one correlation with variance function consisting of a constant
  \(a\) and proportion \(b\):
  \(\Var(\varepsilon_{ij}) = a^2 + b^2t_{ij}^2\), where \(t_{ij}\) is
  continuous-time from baseline,
\item
  \textbf{``CAR1 Exp.''}: continuous-time autoregressive order one
  correlation with
  \(\Var(\varepsilon_{ij}) = \sigma^2 \exp(2\delta t_{ij})\),
\item
  \textbf{``Random Intercept''}: participant-specific random intercepts
  with independent and identically distributed Gaussian residuals, and
\item
  \textbf{``Random Slope''}: participant-specific random intercepts and
  slopes relative to \(t_{ij}\) with independent and identically
  distributed Gaussian residuals.
\end{itemize}

All of the above variance-covariance assumptions and their estimation
are detailed in Pinheiro and Bates\cite{pinheiro2006mixed}. Non-linear models are fit by
maximum likelihood estimation using the \emph{nlme} package \citep{nlme}
in R \citep{R}. Linear models models with unstructured
variance-covariance are fit with the \emph{lme4} package \citep{lme4},
and other covariance structures are fit with \emph{gls} function in the
\emph{nlme} package. The \emph{emmeans} package \citep{emmeans} is used
for estimating the means and contrasts of linear models over time with
Sattherwaite degrees of freedom, while parametric resampling of fixed
effects (based on their estimated mean and covariance) is used for the
nonlinear models. The \emph{ggplot2} package is used for plotting
\citep{ggplot2}. Sample code for the simulation study is provided in the
Appendix.

\hypertarget{model-demonstrations}{%
\section{Model demonstrations}\label{model-demonstrations}}

We demonstrate applications of the models using three completed
Alzheimer's disease (AD) clinical trials. The first is the study of
Donepezil and Vitamin E for Mild Cognitive Impairment (MCI)
\citep{petersen2005vitamin}. For these analyses, we focus on the data
from participants randomized to Donepezil or placebo. The second is a
study of the fyn kinase inhibitor AZD0530 in mild Alzheimer dementia
\citep{van2019effect}. The third is a study of intranasal insulin in MCI
and mild Alzheimer dementia \citep{craft2020safety}. Two intranasal
devices were used in this study because the first was unreliable. We
focus on data from the second device.

None of these studies met their primary endpoint, but Donepezil showed a
benefit in the first months of the MCI trial that seemed to diminish by
the final 36-month time point. The primary analysis approach for the MCI
trial was a Cox proportional hazard model \citep{cox1972} of
time-to-dementia, however the proportional hazard assumption was
violated. For all three trials, we demonstrate the modeling approaches
using the Alzheimer's Disease Assessment Scale--Cognitive Subscale
(ADAS-Cog) \citep{mohs1997development}. For the MCI trial, we see
problems for the proportional treatment effect assumption on ADAS-Cog as
well.

\begin{figure}[!h]

{\centering \includegraphics[width=\linewidth,]{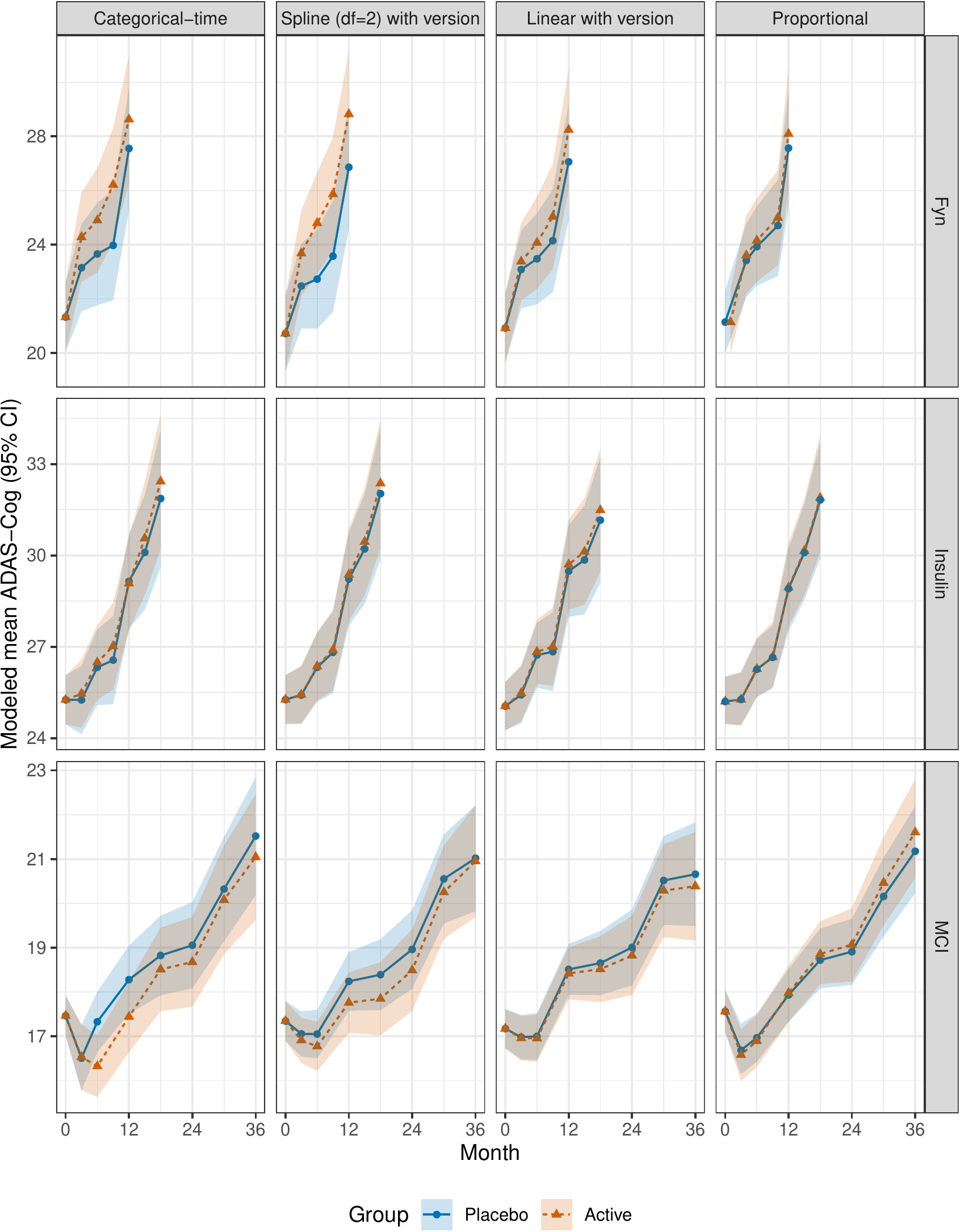} 

}

\caption{Modeled mean ADAS-Cog for each study. All models assume unstructured variance-covariance, except the Proportional model, which assumes random intercepts and slopes. \\ \footnotesize{ADAS-Cog, Alzheimer's Disease Assessment Scale--Cognitive Subscale; MCI, Mild Cognitive Impairment}\label{fig1}}\label{fig:means-adas}
\end{figure}

Figure \ref{fig1} shows the estimated mean trends from each study using
the four approaches assuming unstructured variance-covariance (except
for the proportional model, which assumes random intercepts). We can see
that the spline model with two degrees of freedom (one interior knot)
and a time-varying effect for test version estimates trends that are
very similar to the categorical-time model (left); while the linear and
proportional effect models demonstrate less separation between groups.
The proportional effect model estimates the effect in the MCI trial to
be in the opposite direction of all the other models, perhaps due to the
violation of the proportional effect assumption.

\begin{figure}[!h]

{\centering \includegraphics[width=\linewidth,]{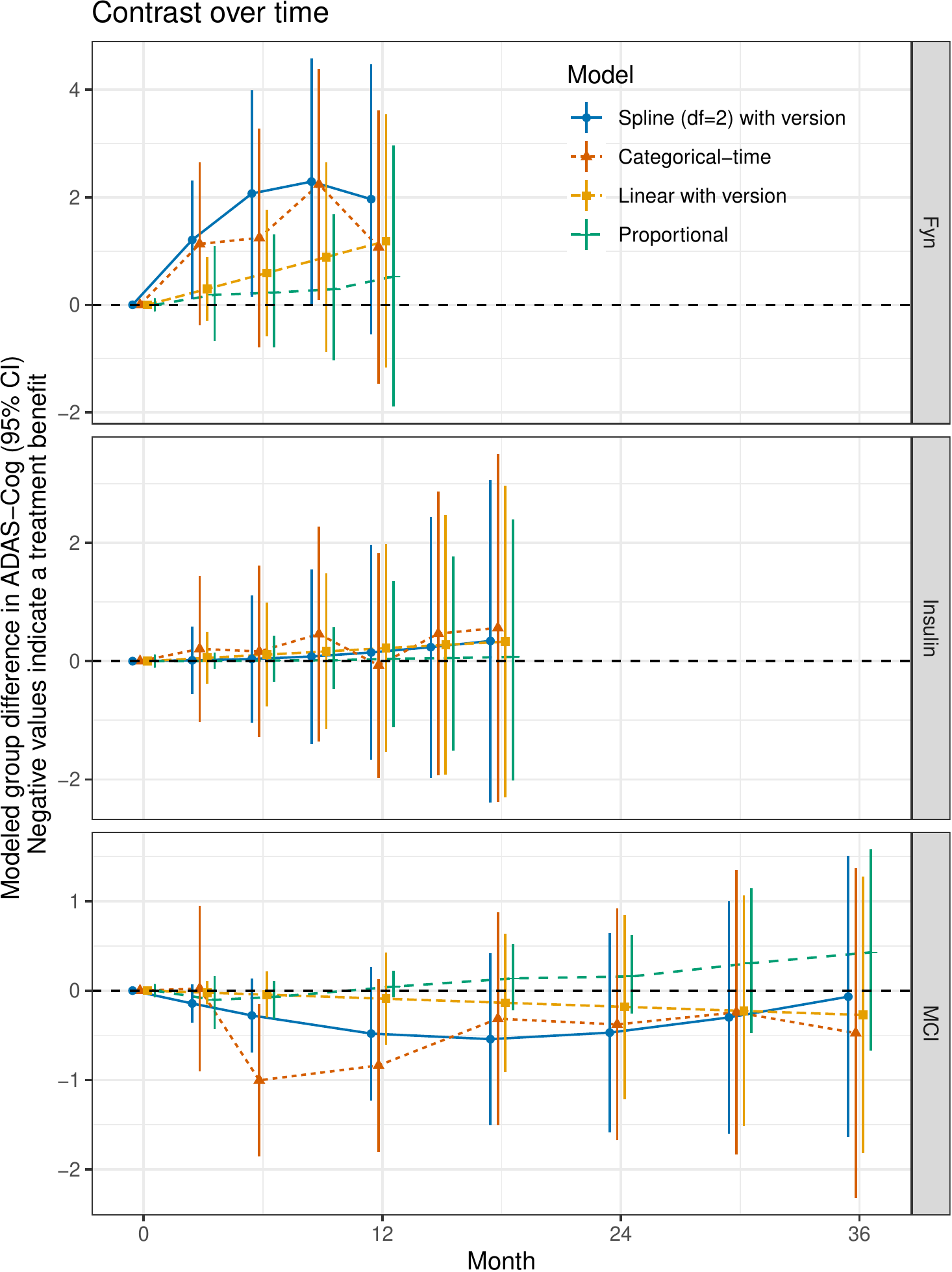} 

}

\caption{Modeled treatment group contrast in ADAS-Cog for each study. All models assume unstructured variance-covariance, except the Proportional model, which assumes random intercepts and slopes. \\ \footnotesize{ADAS-Cog, Alzheimer's Disease Assessment Scale--Cognitive Subscale; MCI, Mild Cognitive Impairment}\label{fig2}}\label{fig:contrasts-adas}
\end{figure}

Figure \ref{fig2} shows the treatment group contrast over time as
estimated by the four approaches in the three trials. In general, the
spline model appears to be a reasonably smoothed version of the trend
estimated using categorical-time; while the linear and proportional
treatment effect models struggle where these assumptions are violated
(i.e.~the Fyn and MCI studies) and treatment effect estimates seem
biased toward the null relative to the other models applied to Fyn and
Insulin. And, again we see the proportional effect model estimating an
effect in the opposite direction when applied to the MCI trial.

\begin{figure}[!h]

{\centering \includegraphics[width=0.90\linewidth,]{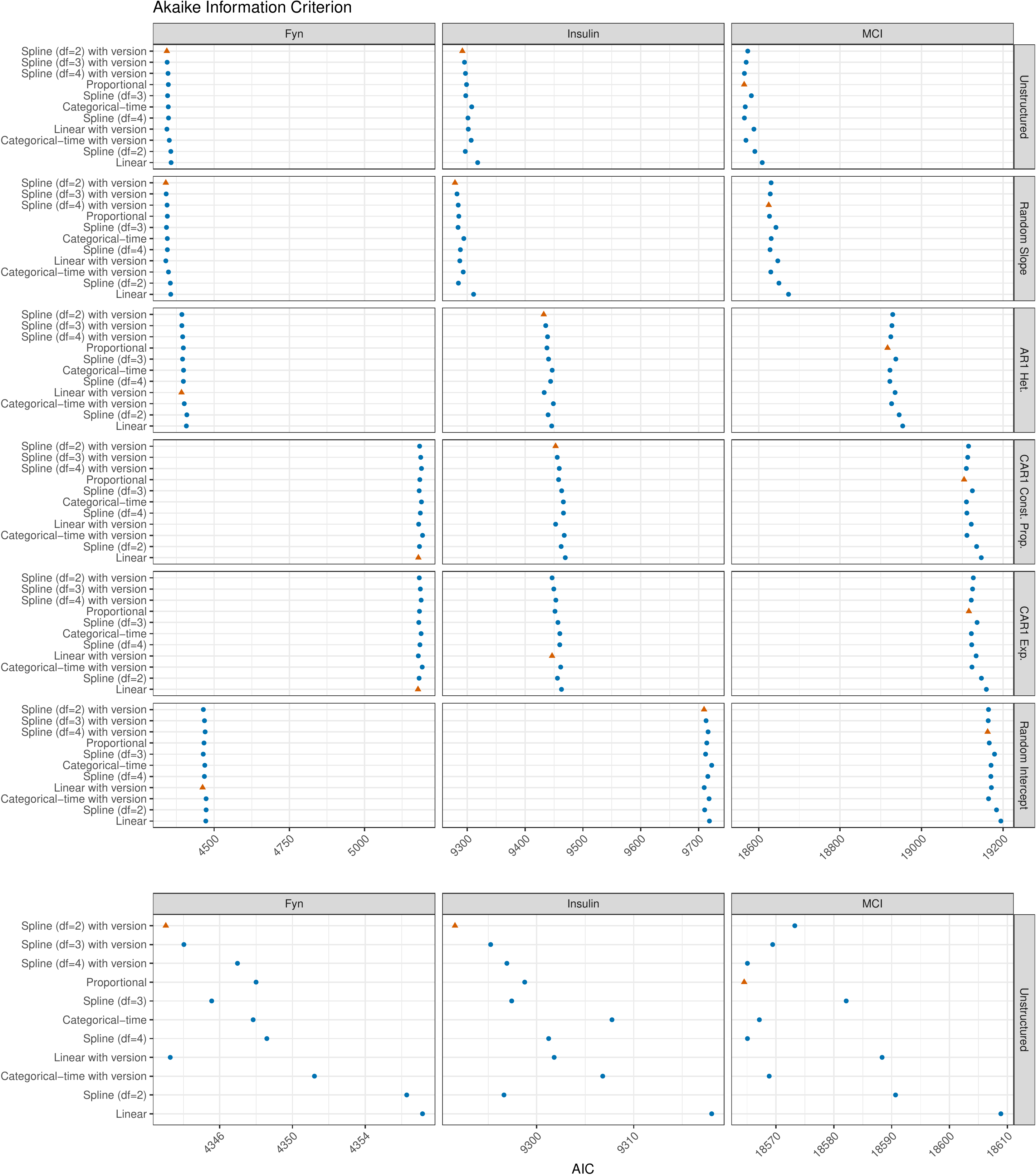} 

}

\caption{AIC for models fit to ADAS-Cog in each study. Models are sorted so that the preferred models appear toward the top, and red triangles indicate the minimum AIC for each study. The Spline (df=2) with version effect is generally preferred. All models here assume residuals are distributed with unstructured heterogeneous variance-covariance.   \\ \footnotesize{AIC, Akaike Information Criterion; ADAS-Cog, Alzheimer's Disease Assessment Scale--Cognitive Subscale; MCI, Mild Cognitive Impairment}\label{fig3}}\label{fig:AIC-plot-adas}
\end{figure}

We also considered Akaike Information Criterion \citep{akaike1974new}
for each model to assess parsimony and predictive value (Figure
\ref{fig3}). In general, the categorical-time unstructured
variance-covariance assumption was preferred even with continuous-time
mean structures, followed closely by the random slopes assumption. The
spline model with two degrees of freedom (one interior knot) and version
effect attained the smallest AIC for the Fyn and Insulin studies. The
proportional model was preferred for MCI, but the proportional effect
assumption is violated in this trial and the resulting treatment effect
estimate is in opposite direction of other models. Considering that the
spline with two degrees of freedom and version effect provide fits that
are similar to categorical-time in Figure \ref{fig1}, we concluded that
two degrees of freedom appear to be sufficient to capture enough of the
general trend to assess the treatment effect.

The AIC comparisons also shows that adding an ADAS-Cog version effect to
the spline model with two degrees of freedom provides a substantial and
consistent reduction in AIC (-13.2, -5, -17.4 AIC points for the Fyn,
Insulin, and MCI trials respectively). Adding a version effect to the
categorical-time model increased AIC or only reduced AIC by a point
(3.4, -1, 1.7 AIC points respectively).

\hypertarget{simulation-studies}{%
\section{Simulation studies}\label{simulation-studies}}

We consider four simulation scenarios. The first two are derived from
the Insulin and MCI studies respectively. Actual participant data
regarding baseline covariates (\emph{APOE} \(\epsilon4\) genotype,
Mini-Mental State Exam {[}MMSE{]}, age) and follow-up (administered test
version {[}A, B, or C{]} and missing data) were used. Within each
simulated trial, group assignments are permuted. Simulated placebo group
outcome data is generated according to a natural cubic spline model fit
to the study data. The model assumes splines with four degrees of
freedom (three interior knots), and covariates for baseline MMSE,
\emph{APOE} \(\epsilon4\) genotype, and age. Residuals are assumed to
have unstructured heterogeneous variance-covariance. The Insulin model
also includes an effect for sex. We do not simulate data from an MMRM
because we want the simulated data to be informed by actual time from
baseline. The choice of four degrees of freedom for the simulation model
is an arbitrary one intended to differ from the candidate natural cubic
spline analysis model with two degrees of freedom.

We simulated clinical trials in four different populations: (1) MCI or
mild Alzheimer dementia, (2) MCI only, (3) Preclinical Alzheimer's
Disease (PAD), and (4) PAD with a COVID19 pandemic disruption in
follow-up. The first two are derived from the Insulin and MCI studies
respectively. Actual participant data from the Insulin and MCI studies
regarding baseline covariates (\emph{APOE} \(\epsilon4\) genotype,
Mini-Mental State Exam {[}MMSE{]}, age) and follow-up (administered test
version {[}A, B, or C{]} and missing data) were used. Within each
simulated trial, group assignments are permuted. Simulated placebo group
outcome data is generated according to a natural cubic spline model fit
to the study data. The model assumes splines with four degrees of
freedom (three interior knots), and covariates for baseline MMSE,
\emph{APOE} \(\epsilon4\) genotype, and age. Residuals are assumed to
have unstructured heterogeneous variance-covariance. The Insulin model
also includes an effect for sex. We do not simulate data from an MMRM
because we want the simulated data to be informed by actual time from
baseline. The choice of four degrees of freedom for the simulation model
is an arbitrary one intended to differ from the candidate natural cubic
spline analysis model with two degrees of freedom.

The treatment benefit for the MCI study is assumed to be linear,
starting at 12 months, reaching a relative benefit of 2.75 ADAS-Cog
points at month 36. This delayed onset of treatment benefit might be
expected from a disease modifying therapy, as opposed a symptomatic
therapy like Donepezil. For the simulation of a trial in MCI or mild
Alzheimer dementia we assume an immediate linear benefit of 4.25
ADAS-Cog points over the course of the 18 month study, which might be
expected from an intervention like insulin which is believed to have the
potential for symptomatic and disease-modifying effects.

The last two simulation studies are based on Preclinical Alzheimer's
Cognitive Composite (PACC) \citep{donohue2014preclinical} data from the
Alzheimer's Disease Neuroimaging Initiative (ADNI; adni.loni.usc.edu)
cognitively normal subjects with evidence of brain amyloid
disregulation, so called Preclinical Alzheimer's Disease (PAD). The ADNI
was launched in 2003 as a public-private partnership, led by Principal
Investigator Michael W. Weiner, MD. The primary goal of ADNI has been to
test whether serial magnetic resonance imaging (MRI), positron emission
tomography (PET), other biological markers, and clinical and
neuropsychological assessment can be combined to measure the progression
of MCI and early AD. The simulated PAD trials include ten study visits
targeted every six months, a target follow-up of (4.5 years). The
simulated observation times are randomly perturbed, centered around the
target time point, according to a normal distribution with standard
deviation of 0.8 months. We also simulate a hypothetical PAD trial
disrupted by COVID19. The COVID interruption is simulated to start
randomly, based on a random sample of the last six visits, and subjects
return to regular visits after a random duration distributed according
to a truncated normal distribution with mean six months, standard
deviation 3 months, and range four to twelve months. Code for this
simulation is provided in the Appendix, including the ADNI-derived model
parameters needed to replicate the simulation. The treatment benefit for
both PAD studies is assumed to be linear relative to the \emph{visit
number}, starting at the fourth visit, and attaining 1.6 PACC points at
the tenth and final visit. This diminishes the treatment effect relative
to time-from-baseline, as might be expected if doses are missed because
of the COVID interruption. Monotone attrition is simulated yielding 30\%
dropout rate by the final visit.

Each of 10,000 simulated trials is analyzed with three mean structures:
a categorical-time model, a natural cubic spline model with two degrees
of freedom and test version effect, and a model assuming a proportional
treatment benefit. The categorical-time model assumed unstructured
heterogeneous variance-covariance. The spline model assumed three
difference variance-covariance assumptions: unstructured heterogeneous,
random slope, and continuous-time autoregressive order one with
exponential variance function of time. The proportional model assumes a
random participant-specific intercept and independent residuals (as
described in \citep{wang2018novel}). The inference for the
categorical-time model is based on the final visit treatment group mean
contrast. The inference for the spline model is based on the final
target time point (i.e.~18 months for Mild AD or MCI, 36 months for MCI
only, and 54 months for both PAD trials). The inference for the
proportional model is based on the estimated proportional treatment
effect (which the model assumes to be constant over time). The
Satterthwaite degrees of freedom were used for all contrasts, except for
the PAD simulations. Due to computational time of the PAD simulations,
Satterthwaite degrees of freedom were only calculated for the first 100
simulated trials, and for subsequent simulations the minimum degrees of
freedom estimated from the first 100 trials was assumed.

Table \ref{tab1} summarizes the simulation results. The spline and
categorical-time models with unstructured variance-covariance provide
reasonable control of Type I error (\textless5.56\%) and the spline
model demonstrated modest but consistent improvement in power in all
scenarios without convergence issues. The spline model improved power
relative to the categorical model by 4.37\% in the Mild AD or MCI
simulation, 9.65\% in the MCI only simulation, 16.07\% in the PAD
simulation, and 13.42\% in the PAD-COVID simulation.

In contrast, the proportional treatment effect model suffered from Type
I error inflation in all scenarios, with Type I error ranging from
27.26\% to 35.28\%. The proportional model was also plagued by model
fitting issues and provided an inferential statistic in only
2407/10000=24.07\% of simulated PAD-COVID trials. Note that the PAD
treatment effects had a delayed onset, a non-proportional effect, which
might explain why the model struggled. The proportional model did not
struggle to converge as much with the Type I error simulations, with no
treatment effect. The Type I error remained inflated for the
proportional model even when models with convergence warnings were
removed (Appendix Table \ref{tab2}).

\begin{table}

\caption{\label{tab:simresults}Simulated power and type I error. Mild AD dementia or MCI and MCI only simulations were based on patient characteristics and model estimates from the Insulin and MCI trials. The PAD simulations are based on patient characteristics and model estimates from ADNI participants. The PAD-COVID simulations add a random interruption in study visits and treatment due to COVID19. The Spline model is based on natural cubic splines with one internal knot and a time-varying test version effect. The number of trials in which a p-value was obtained is indicated by N. \\ \footnotesize{MCI, Mild Cognitive Impairment; AD, Alzheimer's Disease; PAD, Preclinical Alzheimer's Disease; ADNI, Alzheimer's Disease Neuroimaging Initiative; CAR1, continuous-time autoregressive order one correlation with exponential variance function of time; Unstr., unstructured variance-covariance }\label{tab1}}
\centering
\begin{tabular}[t]{llrrrr}
\toprule
\multicolumn{2}{c}{ } & \multicolumn{2}{c}{Power} & \multicolumn{2}{c}{Type I Error} \\
\cmidrule(l{3pt}r{3pt}){3-4} \cmidrule(l{3pt}r{3pt}){5-6}
Study & Model & \% & N & \% & N\\
\cmidrule{1-6}
\midrule
 & Categorical time & 82.29 & 10000 & 5.16 & 10000\\
\cmidrule{2-6}
 & Spline-Unstr. & 86.66 & 10000 & 5.41 & 10000\\
\cmidrule{2-6}
 & Spline-Random slope & 86.10 & 10000 & 5.18 & 10000\\
\cmidrule{2-6}
 & Spline-CAR1 & 82.37 & 10000 & 6.69 & 10000\\
\cmidrule{2-6}
\multirow{-5}{*}{\raggedright\arraybackslash Mild AD or MCI} & Proportional & 95.35 & 9882 & 30.32 & 10000\\
\cmidrule{1-6}
 & Categorical time & 85.32 & 10000 & 5.17 & 10000\\
\cmidrule{2-6}
 & Spline-Unstr. & 94.97 & 10000 & 5.40 & 10000\\
\cmidrule{2-6}
 & Spline-Random slope & 94.48 & 10000 & 5.18 & 10000\\
\cmidrule{2-6}
 & Spline-CAR1 & 84.15 & 10000 & 7.73 & 10000\\
\cmidrule{2-6}
\multirow{-5}{*}{\raggedright\arraybackslash MCI only} & Proportional & 97.08 & 9788 & 27.90 & 10000\\
\cmidrule{1-6}
 & Categorical time & 77.23 & 10000 & 4.98 & 10000\\
\cmidrule{2-6}
 & Spline-Unstr. & 93.30 & 10000 & 5.56 & 10000\\
\cmidrule{2-6}
 & Spline-Random slope & 93.92 & 10000 & 10.73 & 10000\\
\cmidrule{2-6}
 & Spline-CAR1 & 90.81 & 10000 & 8.44 & 10000\\
\cmidrule{2-6}
\multirow{-5}{*}{\raggedright\arraybackslash PAD} & Proportional & 18.52 & 3331 & 27.26 & 9801\\
\cmidrule{1-6}
 & Categorical time & 78.38 & 10000 & 5.30 & 10000\\
\cmidrule{2-6}
 & Spline-Unstr. & 91.80 & 10000 & 5.34 & 10000\\
\cmidrule{2-6}
 & Spline-Random slope & 92.52 & 10000 & 7.06 & 10000\\
\cmidrule{2-6}
 & Spline-CAR1 & 90.96 & 10000 & 8.23 & 10000\\
\cmidrule{2-6}
\multirow{-5}{*}{\raggedright\arraybackslash PAD-COVID} & Proportional & 25.18 & 2407 & 35.28 & 9980\\
\bottomrule
\end{tabular}
\end{table}

\hypertarget{discussion}{%
\section{Discussion}\label{discussion}}

While models with proportional treatment effects have attracted
attention as a purported powerful alternative to MMRM for AD clinical
trials, the model assumptions are often violated and the models appear
to be subject to severe Type I error inflation and convergence problems.
In contrast, spline models have weaker assumptions regarding the nature
of the placebo decline and treatment effect and can generally be fit
with the same software used to fit linear models like MMRM. The weaker
assumptions of the spline model allow a broad possibility of treatment
effects that are not restricted to be linear or proportional relative
the placebo group mean. The spline models also show a modest to
substantial improvement in power (4.37\% to 16.07\%) compared to
categorical-time models and provide reasonable Type I error control. The
increase in power seems to be larger for study designs with more visits,
as one might expect since this increases the complexity of the
categorical-time mean parameterization.

Because the spline model treats time as continuous, it can accommodate
delays in study visits, such as those due to the COVID19 pandemic. With
a categorical-time model, investigators must choose between ignoring
delayed visits, carrying observations back, or creating new visit
categories with depleted observation counts and a mixture of test
versions. The spline model also showed greater power to detect treatment
effects compared to the categorical-time model in the simulated trials
with a COVID19 disruption.

In absence of the COVID19 interruption, the MMRM is targeting the
randomized group mean difference at 4.5 years. An MMRM analysis ignoring
the COVID19 delay in visits is targeting a different estimand: the
randomized group mean difference after nine irregularly spaced follow-up
visits. The test statistic that we explored in the spline analysis, the
treatment group difference at 4.5 years, is therefore more consistent
with the uninterrupted study's original estimand.

A drawback of the spline model is the need to specify the number and
location of interior knots. In these analyses, we followed the default
software setting, which equally spaces the knots according to the
quantiles of observation times. Of note, the number of knots is
understood to be more important than their location, and using quantiles
for knot locations is a broadly recommended approach
\citep{frank2015regression}. The natural cubic spline with one interior
knot at the median observation time seems sufficient in the trials and
datasets that we explored. Conceptually, one interior knot captures
quadratic-like trends with one inflection point for the placebo group
and another for the relative treatment difference. We believe this
degree of flexibility is sufficient to capture the expected trends in
Alzheimer's clinical trials. More exploration might be required in other
applications to ensure a suitable choice of knot locations. Prior
clinical trials or natural history studies can be used to guide this
decision, which should be pre-specified in analytic plans.

\begin{figure}[!h]

{\centering \includegraphics[width=\linewidth,]{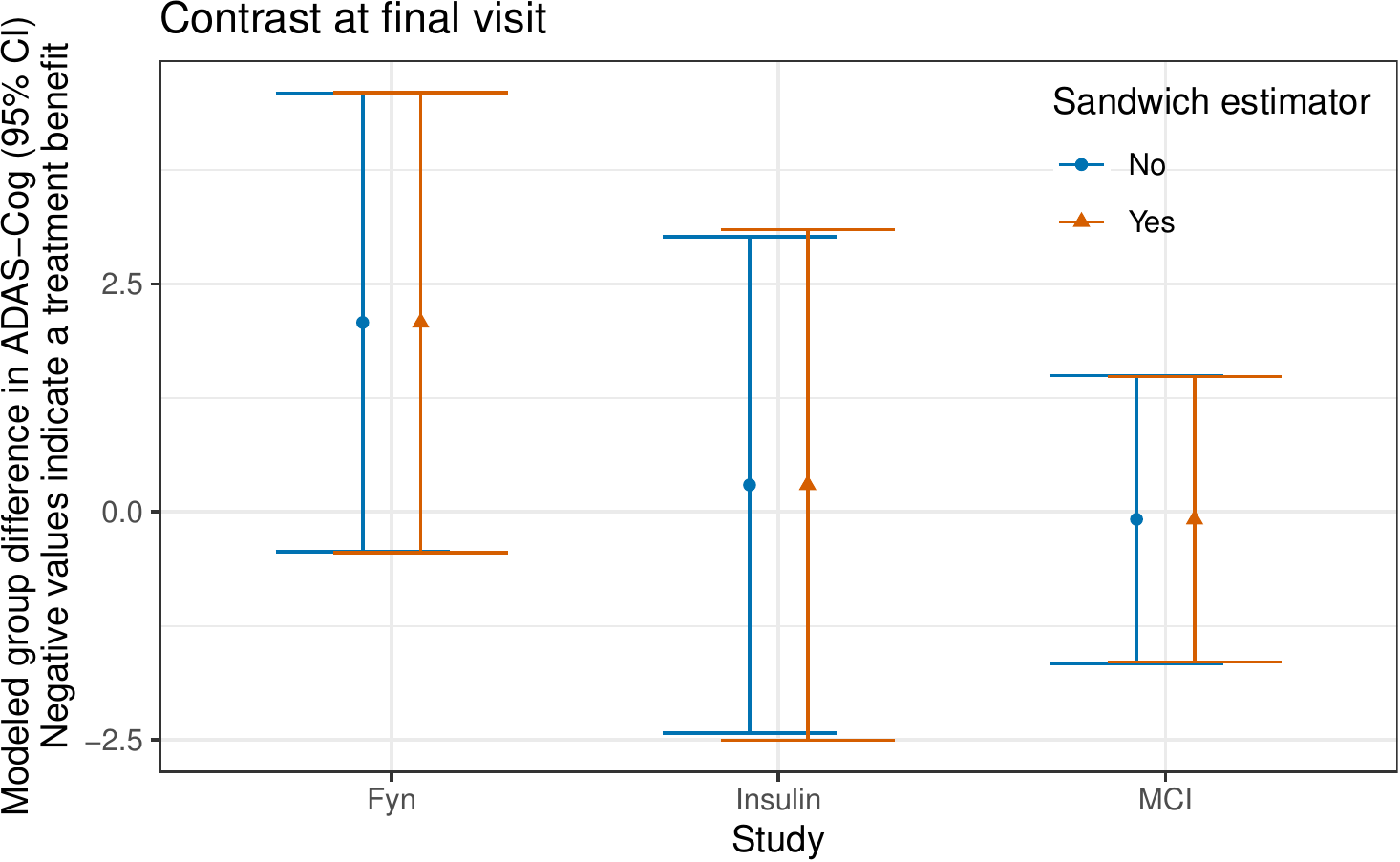} 

}

\caption{Modeled treatment group contrast in ADAS-Cog for each study at the final time with 95\% confidence intervals derived with and without the sandwich estimator of standard error. All estimates are derived from models assuming a natural cubic spline and unstructured variance-covariance. \\ \footnotesize{ADAS-Cog, Alzheimer's Disease Assessment Scale--Cognitive Subscale; MCI, Mild Cognitive Impairment}\label{fig4}}\label{fig:sandwich}
\end{figure}

While unconventional, it appears the combination of a categorical-time
covariance structures and continuous-time mean structure is generally
preferred and supported by AIC in the trials analyzed. We hypothesize
that this is because of the categorical nature of the Alzheimer's
clinical trial visit schedule in which study participants undergo the
same sequence of visits and cognitive test versions. The simulation
studies demonstrate better Type I error control for the spline models
with unstructured variance-covariance (\(\leq\) 5.56\%) relative to
random slope (which had Type I error as large as 10.73\%) and
continuous-time autoregressive order one with exponential variance
function of time (which had Type I error as large as 8.23\%). Our
simulations included some exploration of misspecification of the
variance-covariance, but only when the true model was assumed to have an
unstructured variance-covariance. Sandwich estimators may also be used
to provide standard errors that are robust to misspecification of the
variance-covariance \citep{mccaffrey2003bias, clubSandwich},
however we found this to have negligible effect in the trials analyzed
(Figure \ref{fig4}).

A limitation of this work is that we have focused exclusively on
estimation of the treatment policy estimand using all available
observations submitted to maximum likelihood estimation. These estimates
should be robust and unbiased assuming data are missing at random,
though missing not at random processes were not explored.
Polverejan et al.\citep{polverejan2020aligning} discuss some alternative estimands and
estimation procedure based on multiple imputation for Alzheimer's
clinical trials in the face of data missing not at random relative to
treatment discontinuation and initiation of other therapies. They
demonstrate that maximum likelihood estimators could be biased when on-
and off-treatment mean trajectories differ.

However, we have found that the intuition underlying some hypothetical
estimands and their estimation approaches can be problematic. For
example, intuition might lead one to believe that mean performance on a
cognitive measure would be improved after the initiation of symptomatic
medication for Alzheimer's compared to individuals not initiating
symptomatic medication. And therefore, we might choose to ignore
observations after symptomatic medication and target a hypothetical
estimand for the experimental treatment benefit in absence of the
symptomatic medication. Contrary to this intuition, in individuals with
mild cognitive impairment, allowing observations after the initiation of
symptomatic medication results in \emph{worse} cognitive trajectories than 
censoring those observations \citep{donohue2020initiation}. This is
likely because individuals are prescribed medication because they are
declining, and the benefit conferred by the medication is not enough to
improve that decline above and beyond those who are not prescribed
medication. Therefore we prefer targeting a treatment policy estimand
that is estimated using as much post-randomization data as possible
(regardless of intercurrent therapy) submitted to an appropriate model
under the assumption of data missing at random, and assess the
sensitivity to data missing not at random with a delta method tipping
point analysis \citep{rubin1977formalizing}. We demonstrate how this can
be done using two-level imputation models with the \texttt{mice}
\texttt{R} package \citep{mice2011} in an analysis vignette in the
Appendix. This strategy can also be adapted to use the NCS model to
target the estimands explored by Polverejan, et al.\citep{polverejan2020aligning} This
would require perturbing imputations according to intercurrent events,
rather than simply perturbing all imputations in the active group by the
same amount, as in the tipping point approach.

Overall, the natural cubic spline framework exhibits several advantages
and very few, if any, disadvantages compared to MMRM or models that
assume proportional treatment effects. While the categorical-time MMRM
is virtually assumption-free with regard to the temporal mean trend, the
framework cannot accommodate delays in study visits which makes it
incompatible in practice with intention to treat analyses. The natural
cubic spline model does make assumptions about the temporal mean trend,
but these assumptions seem to reasonably capture group trends. And
adding cognitive test version effects to the spline model results in
very similar estimates as MMRM. Furthermore, imposing the spline
assumptions allows data from delayed visits to be naturally
incorporated, improves power relative to MMRM, and maintains Type I
error control. In contrast, the proportional treatment effect assumption
is too strong and has been violated in at least two AD trials. The
proportional treatment effect models are also challenging to fit and
exhibit unacceptable Type I error.

\clearpage

\hypertarget{acknowledgements}{%
\section{Acknowledgements}\label{acknowledgements}}

Data collection and sharing for this project was funded by the
Alzheimer's Disease Neuroimaging Initiative (ADNI) (National Institutes
of Health Grant U01 AG024904). ADNI is funded by the National Institute
on Aging, the National Institute of Biomedical Imaging and
Bioengineering, and through generous contributions from the following:
AbbVie, Alzheimer's Association; Alzheimer's Drug Discovery Foundation;
Araclon Biotech; BioClinica, Inc.; Biogen; Bristol-Myers Squibb Company;
CereSpir, Inc.; Cogstate; Eisai Inc.; Elan Pharmaceuticals, Inc.; Eli
Lilly and Company; EuroImmun; F. Hoffmann-La Roche Ltd and its
affiliated company Genentech, Inc.; Fujirebio; GE Healthcare; IXICO
Ltd.; Janssen Alzheimer Immunotherapy Research \& Development, LLC.;
Johnson \& Johnson Pharmaceutical Research \& Development LLC.;
Lumosity; Lundbeck; Merck \& Co., Inc.; Meso Scale Diagnostics, LLC.;
NeuroRx Research; Neurotrack Technologies; Novartis Pharmaceuticals
Corporation; Pfizer Inc.; Piramal Imaging; Servier; Takeda
Pharmaceutical Company; and Transition Therapeutics. The Canadian
Institutes of Health Research is providing funds to support ADNI
clinical sites in Canada. Private sector contributions are facilitated
by the Foundation for the National Institutes of Health (www.fnih.org).
The grantee organization is the Northern California Institute for
Research and Education, and the study is coordinated by the Alzheimer's
Therapeutic Research Institute at the University of Southern California.
ADNI data are disseminated by the Laboratory for Neuro Imaging at the
University of Southern California.

The FYN study (ClinicalTrials.gov identifier: NCT02167256) was supported
by grant UH3 TR000967 (Drs Strittmatter, van Dyck, and Nygaard) from the
National Center for Advancing Translational Sciences and grants P50
AG047270 (Dr Strittmatter), P30 AG19610 (Dr Reiman), and R01 AG031581
(Dr Reiman) from the National Institute on Aging. See van Dyck, et
al.~2019 for the complete FYN study team list and other acknowledgments.
The Study of Nasal Insulin in the Fight Against Forgetfulness
(ClinicalTrials.gov identifier: NCT01767909) led by Dr Suzanne Craft was
supported by NIH grant RF1 AG041845. The Mild Cognitive Impairment Study
(ClinicalTrials.gov Identifier: NCT00000173) was led by Dr Ronald C.
Petersen and supported by NIH grants U19 AG010483 and UO1 AG10483.

\hypertarget{data-availability-statement}{%
\section{DATA AVAILABILITY
STATEMENT}\label{data-availability-statement}}

The ADNI data that support the PAD simulations are available from
\url{adni.loni.usc.edu}. The MCI trial data are available from
\url{https://www.adcs.org/data-sharing/}. The AZD0530 and insulin trial
data may be requested from
\href{mailto:biostat_request@atrihub.io}{\nolinkurl{biostat\_request@atrihub.io}}.

\clearpage

\hypertarget{appendix}{%
\section{Appendix}\label{appendix}}

\hypertarget{akaike-information-criterion}{%
\subsection{Akaike Information
Criterion}\label{akaike-information-criterion}}

\clearpage

\hypertarget{simulation-results-excluding-trials-convergence-warnings}{%
\subsection{Simulation results excluding trials convergence
warnings}\label{simulation-results-excluding-trials-convergence-warnings}}

\begin{table}[!h]

\caption{\label{tab:simresults2}Simulated power and type I error. Insulin and MCI simulations were based on patient characteristics and model estimates from those studies. The PAD simulations are based on patient characteristics and model estimates from ADNI participants. The PAD-COVID simulations add a random interruption in study visits and treatment due to COVID19. The Spline model is based on natural cubic splines with one internal knot and a time-varying test version effect. The number of trials in which a p-value was obtained without any convergence warnings is indicated by N. \\ \footnotesize{MCI, Mild Cognitive Impairment; AD, Alzheimer's Disease; PAD, Preclinical Alzheimer's Disease; ADNI, Alzheimer's Disease Neuroimaging Initiative; CAR1, continuous-time autoregressive order one correlation with exponential variance function of time; Unstr., unstructured variance-covariance}\label{tab2}}
\centering
\begin{tabular}[t]{llrrrr}
\toprule
\multicolumn{2}{c}{ } & \multicolumn{2}{c}{Power} & \multicolumn{2}{c}{Type I Error} \\
\cmidrule(l{3pt}r{3pt}){3-4} \cmidrule(l{3pt}r{3pt}){5-6}
Study & Model & \% & N & \% & N\\
\cmidrule{1-6}
\midrule
 & Categorical time & 82.29 & 10000 & 5.16 & 10000\\
\cmidrule{2-6}
 & Spline-Unstr. & 86.66 & 10000 & 5.41 & 10000\\
\cmidrule{2-6}
 & Spline-Random slope & 86.10 & 10000 & 5.18 & 10000\\
\cmidrule{2-6}
 & Spline-CAR1 & 82.37 & 10000 & 6.69 & 10000\\
\cmidrule{2-6}
\multirow{-5}{*}{\raggedright\arraybackslash Mild AD or MCI} & Proportional & 96.25 & 9789 & 30.32 & 10000\\
\cmidrule{1-6}
 & Categorical time & 85.32 & 10000 & 5.17 & 10000\\
\cmidrule{2-6}
 & Spline-Unstr. & 94.97 & 10000 & 5.40 & 10000\\
\cmidrule{2-6}
 & Spline-Random slope & 94.48 & 10000 & 5.18 & 10000\\
\cmidrule{2-6}
 & Spline-CAR1 & 84.15 & 10000 & 7.73 & 10000\\
\cmidrule{2-6}
\multirow{-5}{*}{\raggedright\arraybackslash MCI only} & Proportional & 98.04 & 9692 & 27.90 & 10000\\
\cmidrule{1-6}
 & Categorical time & 77.23 & 10000 & 4.98 & 10000\\
\cmidrule{2-6}
 & Spline-Unstr. & 93.30 & 10000 & 5.56 & 10000\\
\cmidrule{2-6}
 & Spline-Random slope & 93.92 & 10000 & 10.73 & 10000\\
\cmidrule{2-6}
 & Spline-CAR1 & 90.81 & 10000 & 8.44 & 10000\\
\cmidrule{2-6}
\multirow{-5}{*}{\raggedright\arraybackslash PAD} & Proportional & 43.73 & 1347 & 27.58 & 9478\\
\cmidrule{1-6}
 & Categorical time & 78.38 & 10000 & 5.30 & 10000\\
\cmidrule{2-6}
 & Spline-Unstr. & 91.80 & 10000 & 5.34 & 10000\\
\cmidrule{2-6}
 & Spline-Random slope & 92.52 & 10000 & 7.06 & 10000\\
\cmidrule{2-6}
 & Spline-CAR1 & 90.96 & 10000 & 8.23 & 10000\\
\cmidrule{2-6}
\multirow{-5}{*}{\raggedright\arraybackslash PAD-COVID} & Proportional & 46.76 & 1236 & 35.17 & 9777\\
\bottomrule
\end{tabular}
\end{table}

\clearpage

\hypertarget{sample-simulation-code}{%
\subsection{Sample Simulation Code}\label{sample-simulation-code}}

\scriptsize

\begin{verbatim}

library(tidyverse)
library(nlme)
library(lme4)
library(emmeans)
library(splines)

emm_options(lmerTest.limit = 20000)

N <- 1000
NSIMS <- 10000
MC.CORES <- parallel::detectCores() - 2

results_path <- file.path('results')
if(!dir.exists(results_path)) dir.create(results_path, recursive = TRUE)

C <- matrix(
  c(1, 0.791, 0.625, 0.494, 0.391, 0.309, 0.244, 0.193, 0.153, 0.121, 
    0.791, 1, 0.791, 0.625, 0.494, 0.391, 0.309, 0.244, 0.193, 0.153, 
    0.625, 0.791, 1, 0.791, 0.625, 0.494, 0.391, 0.309, 0.244, 0.193, 
    0.494, 0.625, 0.791, 1, 0.791, 0.625, 0.494, 0.391, 0.309, 0.244, 
    0.391, 0.494, 0.625, 0.791, 1, 0.791, 0.625, 0.494, 0.391, 0.309, 
    0.309, 0.391, 0.494, 0.625, 0.791, 1, 0.791, 0.625, 0.494, 0.391, 
    0.244, 0.309, 0.391, 0.494, 0.625, 0.791, 1, 0.791, 0.625, 0.494, 
    0.193, 0.244, 0.309, 0.391, 0.494, 0.625, 0.791, 1, 0.791, 0.625, 
    0.153, 0.193, 0.244, 0.309, 0.391, 0.494, 0.625, 0.791, 1, 0.791, 
    0.121, 0.153, 0.193, 0.244, 0.309, 0.391, 0.494, 0.625, 0.791, 1),
  nrow = 10, byrow = TRUE)
v <- c(2.934, 3.68, 3.597, 3.465, 3.361, 3.791, 4.008, 4.395, 4.886, 7.042)
Sigma <- diag(v) %*% C %*% diag(v)

# longpower::power.mmrm(
#   N=N,
#   Ra = C,
#   ra = c(1, 0.99, 0.98, 0.97, 0.96, 0.95, 0.90, 0.87, 0.80, 0.70),
#   sigmaa = v[10],
#   power = 0.80
# )

# longpower::power.mmrm.ar1(
#   N = N,
#   rho = 0.790729,
#   ra = c(1, 0.99, 0.98, 0.97, 0.96, 0.95, 0.90, 0.87, 0.80, 0.70),
#   sigmaa = v[10],
#   sig.level = 0.05,
#   power = 0.80
# )

DELTA <- 1.4

# simulation data ----

set.seed(20211128)
visits <- tibble(
  visNo = 1:10,
  M = seq(0, 54, by=6),
  version = rep(c('A', 'B', 'C'), length.out=10)
)

subinfo <- tibble(
  id = 1:N,
  age = rnorm(n=N, mean=0, sd=6),
  edu = sample(c(-10.4, -9.4, -8.4, -7.4, -6.4, -5.4, -4.4, -3.4, -2.4, -1.4, 
    -0.4, 0.6, 1.6, 2.6, 3.6),
    size = N, replace = TRUE,
    prob = c(0.001, 0.001, 0.003, 0.001, 0.004, 0.001, 0.072, 0.036, 0.108, 
      0.042, 0.247, 0.039, 0.234, 0.052, 0.159)),
  APOE4 = sample(c(0,1), size=N, replace = TRUE, prob=c(0.70, 0.30)),
  `COVID delay` = extraDistr::rtnorm(n=N, mean=6, sd=3, a=4, b=12),
  `COVID visit start` = sample(5:10, size=N, replace = TRUE),
  `Last visit` = sample(1:10, size=N, prob = c(rep(0.033, 9), 0.703), replace = TRUE)
)

visinfo <- expand_grid(
    id = 1:N,
    visNo = 1:10
  ) %>%
  group_by(visNo) %>%
  mutate(
    Months_jitter = case_when(
      visNo == 1 ~ 0,
      TRUE ~ rnorm(n=N, sd=0.8)
  ))

dd00 <- subinfo %>%
  left_join(visinfo, by='id') %>%
  left_join(visits, by='visNo') %>%
  mutate(
    Month = case_when(
      visNo < `COVID visit start` ~ M + Months_jitter,
      TRUE ~ `COVID delay` + M + Months_jitter),
    yrs = Month/12) %>%
  filter(visNo <= `Last visit`) %>%
  mutate(M = as.factor(M))

dd00 %>%
  group_by(M) %>%
  dplyr::summarize(n=length(M), mean = mean(Month), sd = sd(Month))

# ggplot(dd00 %>% filter(visNo > 1), aes(x=Month, group=M)) + 
#   geom_density(aes(fill = M), alpha=0.2) +
#   ggplot2::scale_fill_discrete()

wide <- model.matrix(~ id + visNo + version + M, dd00) %>%
  as_tibble() %>%
  select(-`(Intercept)`)

# ADNI-derived spline basis functions ----
adni_ns_mat <- splines::ns(dd00$yrs, 
  knots=c(0.4736482, 1.9657769, 4.0082136), 
  Boundary.knots = c(0.000000, 8.476386))

adni_ns_fun <- lapply(1:4, function(x){
  function(t){
    as.numeric(predict(adni_ns_mat, t)[,x])
  }})

dd0 <- dd00 %>%
  left_join(wide, by=c('id', 'visNo')) %>%
  mutate(
    fixef0 = 0.2800923 +
      adni_ns_fun[[1]](yrs)*0.04380665 +
      adni_ns_fun[[2]](yrs)*-0.4601309 +
      adni_ns_fun[[3]](yrs)*-2.232262 +
      adni_ns_fun[[4]](yrs)*-3.509172 +
      APOE4*-0.172294862 + 
      edu*0.247813736 + 
      age*-0.125623763 + 
      `versionB`*0.126458100 + 
      `versionC`*0.266977394
  )

# model fitting functions ----
FIT_cat <- function(x){
  fit1 <- lmer(Y ~ 
      M6 + M12 + M18 + M24 + M30 + M36 + M42 + M48 + M54 +
     (M6 + M12 + M18 + M24 + M30 + M36 + M42 + M48 + M54):Active + 
      APOE4 + age + (0 + M | id),
    x, control = lmerControl(check.nobs.vs.nRE = "ignore"))

  ref_grid(fit1, 
    at = list(
      M6 = 0,
      M12 = 0,
      M18 = 0,
      M24 = 0,
      M30 = 0,
      M36 = 0,
      M42 = 0,
      M48 = 0,
      M54 = 1,
      Active = as.factor(0:1)),
    data = x, mode = "satterthwaite") %>%
    emmeans(specs = 'Active', by = 'M54', lmerTest.limit=2000) %>%
    pairs(reverse=TRUE) %>%
    as.data.frame() %>%
    mutate(mod = 'cat')
}

FIT_ncs_uns <- function(x){
  fit1 <- lmer(Y ~ 
      I(ns21(Month)) + I(ns22(Month)) +
      (I(ns21(Month)) + I(ns22(Month))):Active + 
      APOE4 + age + version + (0 + M | id),
    x, control = lmerControl(check.nobs.vs.nRE = "ignore"))
  
  ref_grid(fit1, 
    at = list(
      Month = 54,
      Active = as.factor(0:1)),
    data = x, mode = "satterthwaite") %>%
    emmeans(specs = 'Active', by = 'Month', lmerTest.limit=2000) %>%
    pairs(reverse=TRUE) %>%
    as.data.frame() %>%
    mutate(mod = 'ncs-uns')
}

FIT_ncs_ranslp <- function(x){
  fit1 <- lmer(Y ~ 
      I(ns21(Month)) + I(ns22(Month)) +
      (I(ns21(Month)) + I(ns22(Month))):Active + 
      APOE4 + age + version + (Month | id),
    x)
  
  ref_grid(fit1, 
    at = list(
      Month = 54,
      Active = as.factor(0:1)),
    data = x, mode = "satterthwaite") %>%
    emmeans(specs = 'Active', by = 'Month', lmerTest.limit=2000) %>%
    pairs(reverse=TRUE) %>%
    as.data.frame() %>%
    mutate(mod = 'ncs-ranslp')
}

FIT_ncs_CAR1 <- function(x){
  fit1 <- gls(Y ~ 
      I(ns21(Month)) + I(ns22(Month)) +
      (I(ns21(Month)) + I(ns22(Month))):Active + 
      APOE4 + age + version,
    x,
    correlation = corCAR1(form = ~ Month | id),
    weights = varExp(form = ~ Month)
  )
  
  ref_grid(fit1, 
    at = list(
      Month = 54,
      Active = as.factor(0:1)),
    data = x, mode = "satterthwaite") %>%
    emmeans(specs = 'Active', by = 'Month', lmerTest.limit=2000) %>%
    pairs(reverse=TRUE) %>%
    as.data.frame() %>%
    mutate(mod = 'ncs-CAR1-Exp')
}

tryCatch.W.E <- function(expr){
  W <- NULL
  w.handler <- function(w){ # warning handler
    W <<- w
    invokeRestart("muffleWarning")
  }
  list(value = withCallingHandlers(tryCatch(expr, error = function(e) e),
    warning = w.handler),
    warning = W)
}

fun_nl0_raw <- function(
  theta, betaM6, betaM12, betaM18, betaM24, betaM30, betaM36, betaM42, betaM48, betaM54,
  beta0, betaage, betaAPOE4,
  M6, M12, M18, M24, M30, M36, M42, M48, M54,
  age, APOE4,
  active)
{
  (M6*betaM6 + 
      M12*(betaM6 + betaM12) + 
      M18*(betaM6 + betaM12 + betaM18) + 
      M24*(betaM6 + betaM12 + betaM18 + betaM24) + 
      M30*(betaM6 + betaM12 + betaM18 + betaM24 + betaM30) + 
      M36*(betaM6 + betaM12 + betaM18 + betaM24 + betaM30 + betaM36) +
      M42*(betaM6 + betaM12 + betaM18 + betaM24 + betaM30 + betaM36 + betaM42) +
      M48*(betaM6 + betaM12 + betaM18 + betaM24 + betaM30 + betaM36 + betaM42 + betaM48) +
      M54*(betaM6 + betaM12 + betaM18 + betaM24 + betaM30 + betaM36 + betaM42 + betaM48 + betaM54))* 
    exp(theta * active) +
    beta0 + age*betaage + APOE4*betaAPOE4
}

FIT_prop <- function(x){
  fit_nls <- nls(Y ~ fun_nl0_raw(
    theta, betaM6, betaM12, betaM18, betaM24, betaM30, betaM36, betaM42, betaM48, betaM54,
    beta0, betaage, betaAPOE4,
    M6, M12, M18, M24, M30, M36, M42, M48, M54,
    age, APOE4,
    active),
    data = x,
    control = nls.control(minFactor=10e-10, warnOnly=TRUE),
    algorithm='port')
  start1 <- coef(fit_nls)
  fit_raw_prop_int <- tryCatch.W.E(nlme(Y ~ fun_nl0_raw(
    theta, betaM6, betaM12, betaM18, betaM24, betaM30, betaM36, betaM42, betaM48, betaM54,
    beta0, betaage, betaAPOE4,
    M6, M12, M18, M24, M30, M36, M42, M48, M54,
    age, APOE4,
    active) + b0,
    data = x,
    fixed = list(theta ~ 1, 
      beta0 ~ 1, betaM6 ~ 1, betaM12 ~ 1, betaM18 ~ 1, betaM24 ~ 1, betaM30 ~ 1, 
      betaM36 ~ 1, betaM42 ~ 1, betaM48 ~ 1, betaM54 ~ 1,
      betaage ~ 1, betaAPOE4 ~ 1),
    random = list(b0 ~ 1),
    groups = ~ id,
    start = start1,
    control = nlmeControl(returnObject=TRUE)))
  if(class(fit_raw_prop_int[[1]])[1]!='nlme'){
    return(tibble(p.value = NA, mod = 'prop', 
      warning = as.character(fit_raw_prop_int[[1]])))
  }else{
    return(summary(fit_raw_prop_int[[1]])$tTable %>%
      as.data.frame() %>%
      rownames_to_column(var='coef') %>%
      filter(coef == 'theta') %>%
      rename(
        estimate = Value,
        SE = `Std.Error`,
        df = DF,
        t.ratio = `t-value`,
        p.value = `p-value`
      ) %>%
      mutate(mod = 'prop', warning = ifelse(
        is.null(fit_raw_prop_int[[2]]), NA, 
        as.character(fit_raw_prop_int[[2]]))))
  }
}

# Seeds ----
set.seed(20211127)
SEEDS <- sample(0:99999999, size=NSIMS, replace=FALSE)
if(any(duplicated(SEEDS))) stop('Duplicated seeds.')
save(SEEDS, file='SEEDS.rdata')

# check for existing results ----
LF <- list.files(file.path(results_path))
completed <- as.numeric(
  gsub('_', '', 
    gsub('.rdata', '', 
      gsub('sim', '', LF))))
TODO <- setdiff(1:NSIMS, completed)

## spline functions ----
ns21 <- function(t){
  as.numeric(predict(splines::ns(dd0$Month, df=2), t)[,1])
}
ns22 <- function(t){
  as.numeric(predict(splines::ns(dd0$Month, df=2), t)[,2])
}

# run simulations ----
parallel::mclapply(rev(TODO), mc.cores = MC.CORES, FUN = function(i){
  print(i)
  set.seed(SEEDS[i])
  resids_w <- mvtnorm::rmvnorm(N,sigma=Sigma)
  colnames(resids_w) <- 1:ncol(resids_w)
  resids <- resids_w %>%
    as_tibble() %>%
    mutate(id = 1:N, active = sample(0:1, size = N, replace=TRUE)) %>%
    pivot_longer(`1`:`10`, names_to = 'visNo', values_to = 'residual') %>%
    mutate(
      visNo = as.numeric(visNo),
      Active = as.factor(active))
  dd1 <- dd0 %>% 
    left_join(resids, by=c('id', 'visNo')) %>%
    mutate(
      Y_type_1 = fixef0 + residual,
      Y_power = case_when(
        visNo <= 4 ~ Y_type_1,
        TRUE ~ Y_type_1 + active * (visNo - 4) * DELTA / (10 - 4)
      )
    )
  dd1$Y <- dd1$Y_power
  POW = list(
    cat = FIT_cat(dd1),
    ncs_uns = FIT_ncs_uns(dd1),
    ncs_ranslp = FIT_ncs_ranslp(dd1),
    ncs_CAR1 = FIT_ncs_CAR1(dd1),
    prop = FIT_prop(dd1))

  dd1$Y <- dd1$Y_type_1
  TYPEI = list(
    cat = FIT_cat(dd1),
    ncs_uns = FIT_ncs_uns(dd1),
    ncs_ranslp = FIT_ncs_ranslp(dd1),
    ncs_CAR1 = FIT_ncs_CAR1(dd1),
    prop = FIT_prop(dd1))
  res <- bind_rows(
    bind_rows(POW) %>% mutate(scenario='power'), 
    bind_rows(TYPEI) %>% mutate(scenario='type I')) %>%
    mutate(sim = i)
  save(res, file = file.path(results_path, paste0('sim_', i, '.rdata')))
  return(NULL)
})
\end{verbatim}

\clearpage

\hypertarget{sandwich-estimation-of-standard-errors}{%
\subsection{Sandwich estimation of standard
errors}\label{sandwich-estimation-of-standard-errors}}

\normalsize

\bibliography{bibfile}

\end{document}

\maketitle

\hypertarget{introduction}{%
\section{Introduction}\label{introduction}}

We demonstrate how one might impute missing data arising in Alzheimer's
Disease trials using multilevel models as implemented in the Multiple
Imputation by Chained Equations (\texttt{mice}) \texttt{R} package (van
Buuren (2018), van Buuren and Oudshoorn (2011)). We adapt a
demonstrations provided in
\href{https://stefvanbuuren.name/fimd/sec-multioutcome.html}{van Buuren
(2018)} to our simulated trial data, and show how the multiple
imputations generated under the assumption of Missing at Random (MAR)
can be perturbed in a Missing Not at Random (MNAR) tipping point
sensitivity analyses (Rubin 1977) for a Natural Cubic Spline (NCS)
constrained longitudinal data analysis.

\hypertarget{load-required-r-packages-and-simulated-data}{%
\section{\texorpdfstring{Load required \texttt{R} packages and simulated
data}{Load required R packages and simulated data}}\label{load-required-r-packages-and-simulated-data}}

\begin{Shaded}
\begin{Highlighting}[]
\FunctionTok{library}\NormalTok{(tidyverse)}
\FunctionTok{library}\NormalTok{(lme4)}
\FunctionTok{library}\NormalTok{(emmeans)}
\FunctionTok{library}\NormalTok{(splines)}
\FunctionTok{library}\NormalTok{(mice)}
\FunctionTok{library}\NormalTok{(miceadds)}
\FunctionTok{library}\NormalTok{(pan)}

\FunctionTok{emm\_options}\NormalTok{(}\AttributeTok{lmerTest.limit =} \DecValTok{20000}\NormalTok{)}

\CommentTok{\# load data {-}{-}{-}{-}}
\FunctionTok{load}\NormalTok{(}\StringTok{\textquotesingle{}pad\_covid.Rdata\textquotesingle{}}\NormalTok{)}
\end{Highlighting}
\end{Shaded}

\hypertarget{augment-data-with-missing-timepoints}{%
\section{Augment data with missing
timepoints}\label{augment-data-with-missing-timepoints}}

First we augment data, adding rows for missing Preclinical Alzheimer's
Cognitive Composite (PACC) (Donohue et al. (2014)) observations at the
targeted final time point, 54 months, for any participants without an
observation beyond 48 months.

\begin{Shaded}
\begin{Highlighting}[]
\DocumentationTok{\#\# Filter the subjects without PACC after 48 Months {-}{-}{-}{-}}
\NormalTok{dropouts }\OtherTok{\textless{}{-}}\NormalTok{ pad\_covid }\SpecialCharTok{\%\textgreater{}\%}
  \FunctionTok{group\_by}\NormalTok{(id) }\SpecialCharTok{\%\textgreater{}\%}
  \FunctionTok{summarize}\NormalTok{(}\AttributeTok{max\_month =} \FunctionTok{max}\NormalTok{(Month)) }\SpecialCharTok{\%\textgreater{}\%}
  \FunctionTok{filter}\NormalTok{(max\_month }\SpecialCharTok{\textless{}} \DecValTok{54{-}6}\NormalTok{) }\SpecialCharTok{\%\textgreater{}\%}
  \FunctionTok{pull}\NormalTok{(id)}

\DocumentationTok{\#\# Identify final level of categorical M variable {-}{-}{-}{-}}
\NormalTok{M\_levels }\OtherTok{\textless{}{-}} \FunctionTok{unique}\NormalTok{(pad\_covid}\SpecialCharTok{$}\NormalTok{M)}
\NormalTok{M\_54 }\OtherTok{\textless{}{-}}\NormalTok{ M\_levels[M\_levels}\SpecialCharTok{==}\StringTok{\textquotesingle{}54\textquotesingle{}}\NormalTok{]}

\DocumentationTok{\#\# augment data {-}{-}{-}{-}}
\NormalTok{pad\_covid\_augment }\OtherTok{\textless{}{-}} \FunctionTok{filter}\NormalTok{(pad\_covid, id }\SpecialCharTok{\%in\%}\NormalTok{ dropouts) }\SpecialCharTok{\%\textgreater{}\%}
  \FunctionTok{arrange}\NormalTok{(id, Month) }\SpecialCharTok{\%\textgreater{}\%}
  \FunctionTok{filter}\NormalTok{(}\SpecialCharTok{!}\FunctionTok{duplicated}\NormalTok{(id)) }\SpecialCharTok{\%\textgreater{}\%}
  \FunctionTok{mutate}\NormalTok{(}\AttributeTok{Month =} \DecValTok{54}\NormalTok{, }\AttributeTok{M =}\NormalTok{ M\_54, }\AttributeTok{Y =} \ConstantTok{NA}\NormalTok{, }\AttributeTok{version =} \StringTok{\textquotesingle{}A\textquotesingle{}}\NormalTok{) }\SpecialCharTok{\%\textgreater{}\%}
  \FunctionTok{bind\_rows}\NormalTok{(pad\_covid)}
\end{Highlighting}
\end{Shaded}

\hypertarget{explore-multiple-imputation-models}{%
\section{Explore multiple imputation
models}\label{explore-multiple-imputation-models}}

We follow the example provide by van Buuren (2018)
(\url{https://stefvanbuuren.name/fimd/sec-multioutcome.html}) for
multilevel data to explore different MAR imputation schemes, including:

\begin{itemize}
\tightlist
\item
  \texttt{sample}: simple random sampling of observed values
\item
  \texttt{pmm}: predictive mean matching
\item
  \texttt{2l.pan}: two-level normal model using multivariate linear
  mixed model as implemented in the \texttt{pan} package (Joseph L.
  Schafer 2018)
\item
  \texttt{2l.norm}: two-level normal model
\item
  \texttt{2l.lmer}: two-level normal model using linear mixed models
\item
  \texttt{2l.pmm}: two-level normal model using predictive mean matching
\end{itemize}

In the augmented data, only the final target visit, month 54, is missing
by design:

\begin{Shaded}
\begin{Highlighting}[]
\FunctionTok{md.pattern}\NormalTok{(pad\_covid\_augment[,}\FunctionTok{c}\NormalTok{(}\StringTok{\textquotesingle{}M54\textquotesingle{}}\NormalTok{, }\StringTok{\textquotesingle{}Y\textquotesingle{}}\NormalTok{)], }\AttributeTok{plot =} \ConstantTok{TRUE}\NormalTok{)}
\end{Highlighting}
\end{Shaded}

\begin{center}\includegraphics[width=\linewidth,]{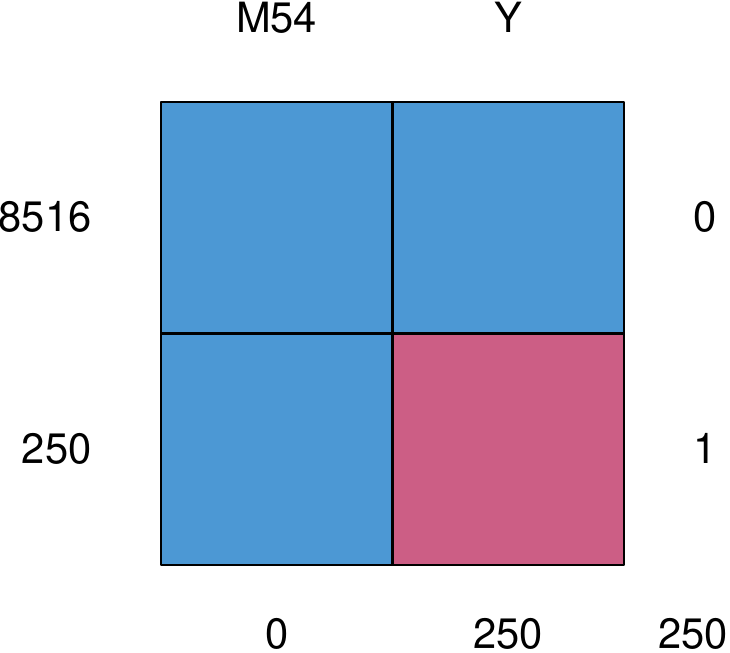} \end{center}

\begin{verbatim}
     M54   Y    
8516   1   1   0
250    1   0   1
       0 250 250
\end{verbatim}

\begin{Shaded}
\begin{Highlighting}[]
\CommentTok{\# select only the columns necessary for imputation:}
\NormalTok{d }\OtherTok{\textless{}{-}}\NormalTok{ pad\_covid\_augment }\SpecialCharTok{\%\textgreater{}\%} \FunctionTok{select}\NormalTok{(id, Month, active, APOE4, age, version, Y)}
\NormalTok{pred }\OtherTok{\textless{}{-}} \FunctionTok{make.predictorMatrix}\NormalTok{(d)}

\CommentTok{\# The code {-}2 in the predictor matrix pred signals that id is the cluster variable}
\NormalTok{pred[, }\StringTok{"id"}\NormalTok{] }\OtherTok{\textless{}{-}} \SpecialCharTok{{-}}\DecValTok{2}
\NormalTok{pred[}\StringTok{"id"}\NormalTok{, }\StringTok{"id"}\NormalTok{] }\OtherTok{\textless{}{-}} \DecValTok{0}

\NormalTok{methods }\OtherTok{\textless{}{-}} \FunctionTok{c}\NormalTok{(}\StringTok{"sample"}\NormalTok{, }\StringTok{"pmm"}\NormalTok{, }\StringTok{"2l.pan"}\NormalTok{, }\StringTok{"2l.norm"}\NormalTok{, }\StringTok{"2l.lmer"}\NormalTok{, }\StringTok{"2l.pmm"}\NormalTok{)}
\NormalTok{result }\OtherTok{\textless{}{-}} \FunctionTok{vector}\NormalTok{(}\StringTok{"list"}\NormalTok{, }\FunctionTok{length}\NormalTok{(methods))}
\FunctionTok{names}\NormalTok{(result) }\OtherTok{\textless{}{-}}\NormalTok{ methods}
\ControlFlowTok{for}\NormalTok{ (meth }\ControlFlowTok{in}\NormalTok{ methods) \{}
  \CommentTok{\# There is only one variable with missing values here, so we do not }
  \CommentTok{\# need to iterate, and can set maxit = 1}
\NormalTok{  result[[meth]] }\OtherTok{\textless{}{-}} \FunctionTok{mice}\NormalTok{(d, }\AttributeTok{predictorMatrix =}\NormalTok{ pred, }\AttributeTok{method =}\NormalTok{ meth,}
    \AttributeTok{m =} \DecValTok{10}\NormalTok{, }\AttributeTok{maxit =} \DecValTok{1}\NormalTok{,}
    \AttributeTok{print =} \ConstantTok{FALSE}\NormalTok{, }\AttributeTok{seed =} \DecValTok{82828}\NormalTok{)}
\NormalTok{\}}
\end{Highlighting}
\end{Shaded}

\hypertarget{visualize-imputations}{%
\section{Visualize imputations}\label{visualize-imputations}}

\begin{Shaded}
\begin{Highlighting}[]
\NormalTok{pad\_covid }\SpecialCharTok{\%\textgreater{}\%}
  \FunctionTok{group\_by}\NormalTok{(id) }\SpecialCharTok{\%\textgreater{}\%}
  \FunctionTok{summarize}\NormalTok{(}\AttributeTok{SD =} \FunctionTok{sd}\NormalTok{(Y)) }\SpecialCharTok{\%\textgreater{}\%}
  \FunctionTok{ggplot}\NormalTok{(}\FunctionTok{aes}\NormalTok{(}\AttributeTok{x=}\NormalTok{SD)) }\SpecialCharTok{+}
  \FunctionTok{geom\_histogram}\NormalTok{() }\SpecialCharTok{+}
  \FunctionTok{xlab}\NormalTok{(}\StringTok{\textquotesingle{}SD(PACC) per subject ID\textquotesingle{}}\NormalTok{)}
\end{Highlighting}
\end{Shaded}

\begin{figure}[!h]

{\centering \includegraphics[width=\linewidth,]{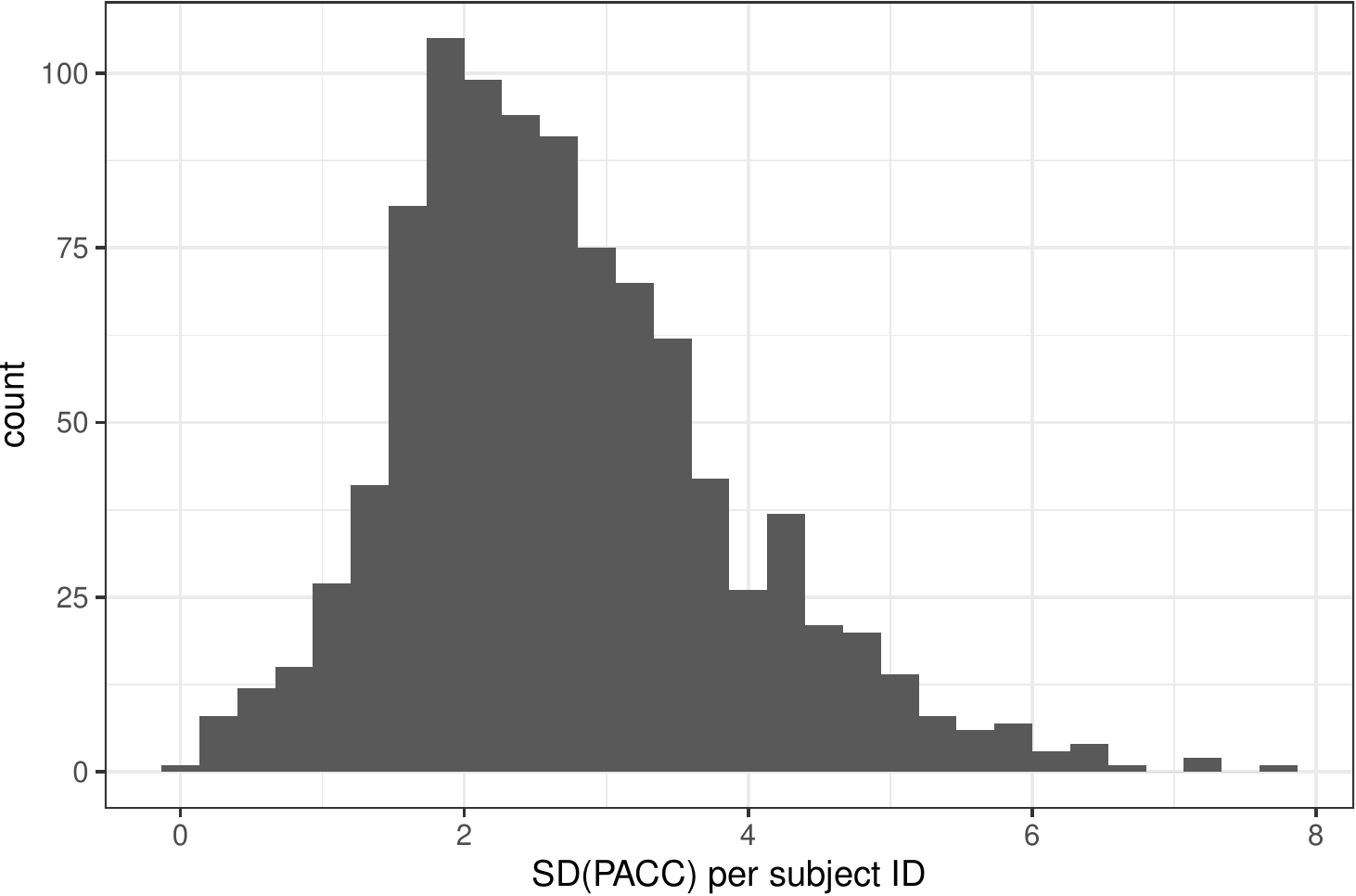} 

}

\caption{Distribution of standard deviations of PACC scores per participant. The standard deviations within participants varies betweeen zero and almost eight PACC points.}\label{fig:histogram}
\end{figure}

\begin{Shaded}
\begin{Highlighting}[]
\NormalTok{result2 }\OtherTok{\textless{}{-}} \FunctionTok{lapply}\NormalTok{(}\FunctionTok{names}\NormalTok{(result), }\ControlFlowTok{function}\NormalTok{(meth)\{}
  \FunctionTok{complete}\NormalTok{(result[[meth]], }\StringTok{"long"}\NormalTok{, }\AttributeTok{include=}\ConstantTok{FALSE}\NormalTok{) }\SpecialCharTok{\%\textgreater{}\%} 
    \FunctionTok{as.data.frame}\NormalTok{() }\SpecialCharTok{\%\textgreater{}\%}
    \FunctionTok{mutate}\NormalTok{(}\AttributeTok{Method =}\NormalTok{ meth)}
\NormalTok{\}) }\SpecialCharTok{\%\textgreater{}\%} \FunctionTok{bind\_rows}\NormalTok{() }\SpecialCharTok{\%\textgreater{}\%}
  \FunctionTok{mutate}\NormalTok{(}
    \AttributeTok{Observed =} \FunctionTok{case\_when}\NormalTok{(id }\SpecialCharTok{\%in\%}\NormalTok{ dropouts }\SpecialCharTok{\&}\NormalTok{ Month }\SpecialCharTok{==} \DecValTok{54} \SpecialCharTok{\textasciitilde{}} \StringTok{\textquotesingle{}Imputed\textquotesingle{}}\NormalTok{,}
      \ConstantTok{TRUE} \SpecialCharTok{\textasciitilde{}} \StringTok{\textquotesingle{}Observed\textquotesingle{}}\NormalTok{),}
    \AttributeTok{.imp =} \FunctionTok{as.numeric}\NormalTok{(.imp),}
    \AttributeTok{.imp =} \FunctionTok{case\_when}\NormalTok{(Observed }\SpecialCharTok{==} \StringTok{\textquotesingle{}Observed\textquotesingle{}} \SpecialCharTok{\textasciitilde{}} \DecValTok{99}\NormalTok{, }\ConstantTok{TRUE} \SpecialCharTok{\textasciitilde{}}\NormalTok{ .imp))}

\NormalTok{result2  }\SpecialCharTok{\%\textgreater{}\%}
  \FunctionTok{filter}\NormalTok{(Month }\SpecialCharTok{\textless{}=} \DecValTok{60} \SpecialCharTok{\&}\NormalTok{ Month }\SpecialCharTok{\textgreater{}=} \DecValTok{48}\NormalTok{) }\SpecialCharTok{\%\textgreater{}\%}
  \FunctionTok{filter}\NormalTok{(Observed }\SpecialCharTok{==} \StringTok{\textquotesingle{}Imputed\textquotesingle{}} \SpecialCharTok{|}\NormalTok{ Method }\SpecialCharTok{==} \StringTok{\textquotesingle{}sample\textquotesingle{}}\NormalTok{) }\SpecialCharTok{\%\textgreater{}\%}
  \FunctionTok{mutate}\NormalTok{(}
    \AttributeTok{Method =} \FunctionTok{case\_when}\NormalTok{(Observed }\SpecialCharTok{==} \StringTok{\textquotesingle{}Observed\textquotesingle{}} \SpecialCharTok{\textasciitilde{}} \StringTok{\textquotesingle{}Observed\textquotesingle{}}\NormalTok{, }\ConstantTok{TRUE} \SpecialCharTok{\textasciitilde{}}\NormalTok{ Method)) }\SpecialCharTok{\%\textgreater{}\%}
  \FunctionTok{ggplot}\NormalTok{(}\FunctionTok{aes}\NormalTok{(}\AttributeTok{x=}\NormalTok{Y)) }\SpecialCharTok{+}
  \FunctionTok{geom\_boxplot}\NormalTok{(}\FunctionTok{aes}\NormalTok{(}\AttributeTok{color=}\NormalTok{Observed)) }\SpecialCharTok{+}
  \FunctionTok{facet\_grid}\NormalTok{(}\AttributeTok{rows =} \FunctionTok{vars}\NormalTok{(Method)) }\SpecialCharTok{+}
  \FunctionTok{xlab}\NormalTok{(}\StringTok{\textquotesingle{}Imputed PACC near 54 months\textquotesingle{}}\NormalTok{) }\SpecialCharTok{+}
  \FunctionTok{theme}\NormalTok{(}\AttributeTok{legend.position =} \StringTok{\textquotesingle{}none\textquotesingle{}}\NormalTok{)}
\end{Highlighting}
\end{Shaded}

\begin{figure}[!h]

{\centering \includegraphics[width=\linewidth,]{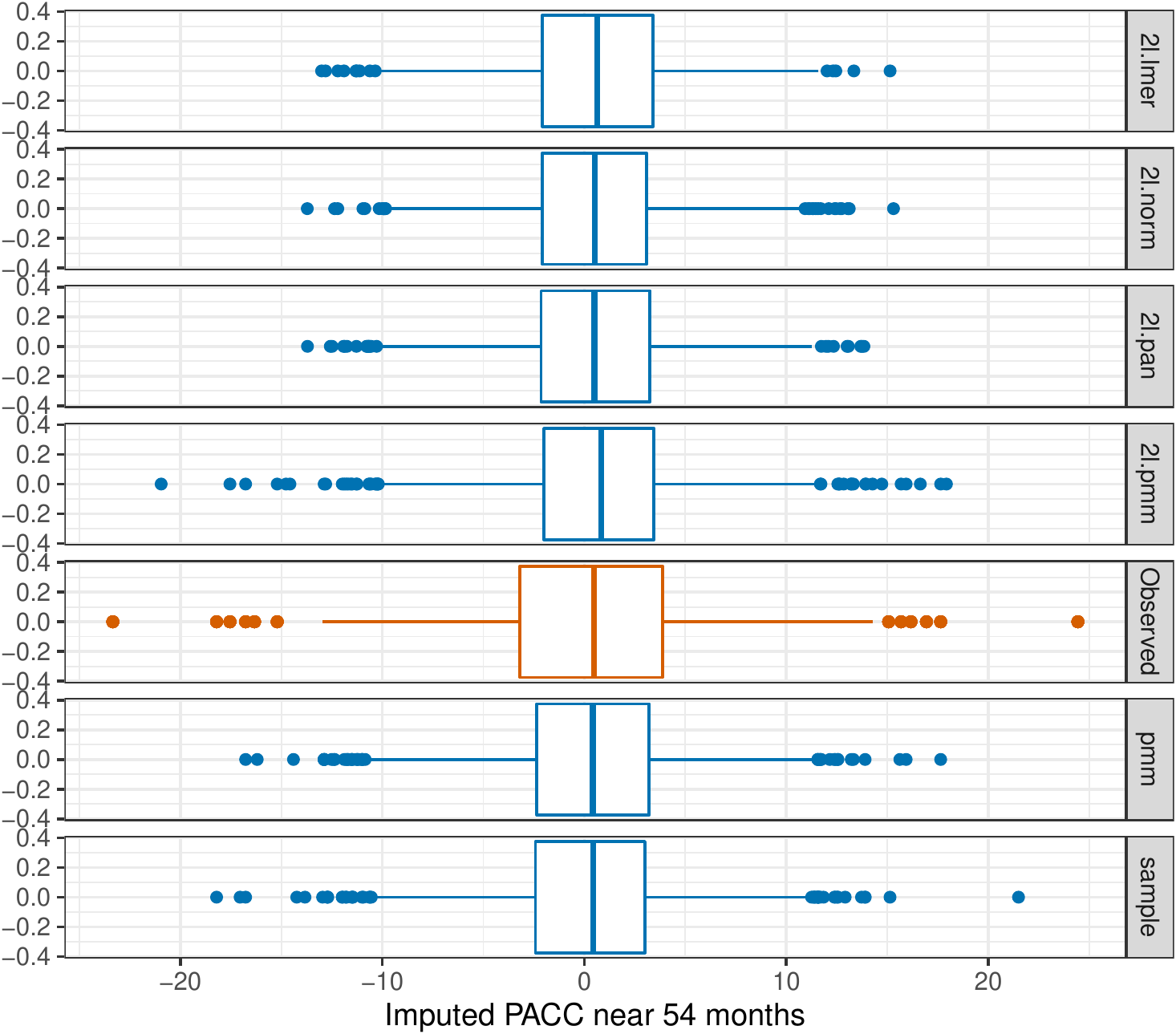} 

}

\caption{Box plots comparing the distribution of the observed data within 54 $\pm$ 6 months (red), and the imputed data (blue) under various methods. Imputations using `2l.pmm` seem to best match the range of observed data.}\label{fig:boxplot}
\end{figure}

\begin{Shaded}
\begin{Highlighting}[]
\NormalTok{result2  }\SpecialCharTok{\%\textgreater{}\%}
  \FunctionTok{filter}\NormalTok{(Month }\SpecialCharTok{\textless{}=} \DecValTok{60} \SpecialCharTok{\&}\NormalTok{ Month }\SpecialCharTok{\textgreater{}=} \DecValTok{48}\NormalTok{) }\SpecialCharTok{\%\textgreater{}\%}
  \FunctionTok{ggplot}\NormalTok{(}\FunctionTok{aes}\NormalTok{(}\AttributeTok{x=}\NormalTok{Y)) }\SpecialCharTok{+}
  \FunctionTok{geom\_density}\NormalTok{(}\FunctionTok{aes}\NormalTok{(}\AttributeTok{color=}\NormalTok{Observed, }\AttributeTok{group=}\NormalTok{.imp)) }\SpecialCharTok{+}
  \FunctionTok{facet\_grid}\NormalTok{(}\AttributeTok{rows =} \FunctionTok{vars}\NormalTok{(Method)) }\SpecialCharTok{+}
  \FunctionTok{xlab}\NormalTok{(}\StringTok{\textquotesingle{}Imputed PACC near 54 months\textquotesingle{}}\NormalTok{) }\SpecialCharTok{+}
  \FunctionTok{theme}\NormalTok{(}\AttributeTok{legend.position =} \StringTok{\textquotesingle{}none\textquotesingle{}}\NormalTok{)}
\end{Highlighting}
\end{Shaded}

\begin{figure}[!h]

{\centering \includegraphics[width=\linewidth,]{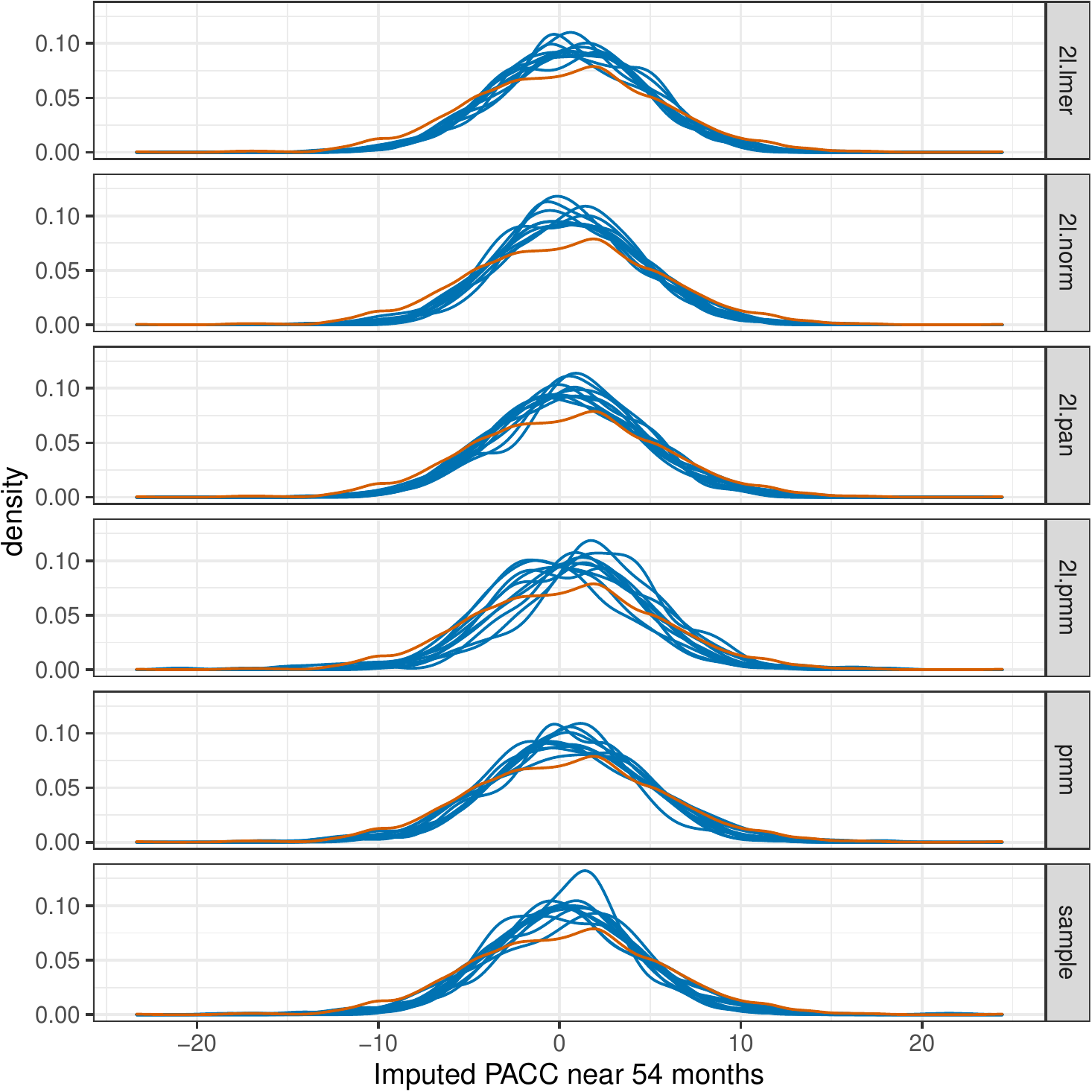} 

}

\caption{Density plots comparing the distribution of the observed data within 54 $\pm$ 6 months (red), and the imputed data (blue) under various methods across the ten imputations.}\label{fig:densities}
\end{figure}

\hypertarget{pool-2l.pmm-imputations}{%
\section{\texorpdfstring{Pool \texttt{2l.pmm}
imputations}{Pool 2l.pmm imputations}}\label{pool-2l.pmm-imputations}}

\begin{Shaded}
\begin{Highlighting}[]
\DocumentationTok{\#\# spline basis expansion functions for NCS model {-}{-}{-}{-}}
\NormalTok{ns21 }\OtherTok{\textless{}{-}} \ControlFlowTok{function}\NormalTok{(t)\{}
  \FunctionTok{as.numeric}\NormalTok{(}\FunctionTok{predict}\NormalTok{(splines}\SpecialCharTok{::}\FunctionTok{ns}\NormalTok{(pad\_covid}\SpecialCharTok{$}\NormalTok{Month, }\AttributeTok{df=}\DecValTok{2}\NormalTok{), t)[,}\DecValTok{1}\NormalTok{])}
\NormalTok{\}}
\NormalTok{ns22 }\OtherTok{\textless{}{-}} \ControlFlowTok{function}\NormalTok{(t)\{}
  \FunctionTok{as.numeric}\NormalTok{(}\FunctionTok{predict}\NormalTok{(splines}\SpecialCharTok{::}\FunctionTok{ns}\NormalTok{(pad\_covid}\SpecialCharTok{$}\NormalTok{Month, }\AttributeTok{df=}\DecValTok{2}\NormalTok{), t)[,}\DecValTok{2}\NormalTok{])}
\NormalTok{\}}

\CommentTok{\# lmer fit to observed data {-}{-}{-}{-}}

\NormalTok{fit\_observed }\OtherTok{\textless{}{-}} \FunctionTok{lmer}\NormalTok{(Y }\SpecialCharTok{\textasciitilde{}} 
    \FunctionTok{I}\NormalTok{(}\FunctionTok{ns21}\NormalTok{(Month)) }\SpecialCharTok{+} \FunctionTok{I}\NormalTok{(}\FunctionTok{ns22}\NormalTok{(Month)) }\SpecialCharTok{+}
\NormalTok{    (}\FunctionTok{I}\NormalTok{(}\FunctionTok{ns21}\NormalTok{(Month)) }\SpecialCharTok{+} \FunctionTok{I}\NormalTok{(}\FunctionTok{ns22}\NormalTok{(Month)))}\SpecialCharTok{:}\NormalTok{active }\SpecialCharTok{+} 
\NormalTok{    APOE4 }\SpecialCharTok{+}\NormalTok{ age }\SpecialCharTok{+}\NormalTok{ version }\SpecialCharTok{+}\NormalTok{ (}\DecValTok{1} \SpecialCharTok{|}\NormalTok{ id),}
\NormalTok{  pad\_covid)}

\NormalTok{rg\_observed }\OtherTok{\textless{}{-}} \FunctionTok{ref\_grid}\NormalTok{(fit\_observed, }
  \AttributeTok{at =} \FunctionTok{list}\NormalTok{(}
    \AttributeTok{Month =} \DecValTok{54}\NormalTok{,}
    \AttributeTok{active =} \FunctionTok{c}\NormalTok{(}\DecValTok{0}\NormalTok{,}\DecValTok{1}\NormalTok{)),}
  \AttributeTok{data =}\NormalTok{ pad\_covid, }\AttributeTok{mode =} \StringTok{"satterthwaite"}\NormalTok{)}

\NormalTok{contrast\_observed }\OtherTok{\textless{}{-}} \FunctionTok{emmeans}\NormalTok{(rg\_observed, }\AttributeTok{specs =} \StringTok{\textquotesingle{}active\textquotesingle{}}\NormalTok{, }
    \AttributeTok{by =} \StringTok{\textquotesingle{}Month\textquotesingle{}}\NormalTok{, }\AttributeTok{lmerTest.limit=}\DecValTok{2000}\NormalTok{) }\SpecialCharTok{\%\textgreater{}\%}
  \FunctionTok{pairs}\NormalTok{(}\AttributeTok{reverse=}\ConstantTok{TRUE}\NormalTok{)}

\CommentTok{\# lmer fits to data with imputations {-}{-}{-}{-}}

\NormalTok{fit\_mi }\OtherTok{\textless{}{-}} \FunctionTok{with}\NormalTok{(result}\SpecialCharTok{$}\StringTok{\textasciigrave{}}\AttributeTok{2l.pmm}\StringTok{\textasciigrave{}}\NormalTok{, }\FunctionTok{lmer}\NormalTok{(Y }\SpecialCharTok{\textasciitilde{}} 
    \FunctionTok{I}\NormalTok{(}\FunctionTok{ns21}\NormalTok{(Month)) }\SpecialCharTok{+} \FunctionTok{I}\NormalTok{(}\FunctionTok{ns22}\NormalTok{(Month)) }\SpecialCharTok{+}
\NormalTok{    (}\FunctionTok{I}\NormalTok{(}\FunctionTok{ns21}\NormalTok{(Month)) }\SpecialCharTok{+} \FunctionTok{I}\NormalTok{(}\FunctionTok{ns22}\NormalTok{(Month)))}\SpecialCharTok{:}\NormalTok{active }\SpecialCharTok{+} 
\NormalTok{    APOE4 }\SpecialCharTok{+}\NormalTok{ age }\SpecialCharTok{+}\NormalTok{ version }\SpecialCharTok{+}\NormalTok{ (}\DecValTok{1} \SpecialCharTok{|}\NormalTok{ id)))}

\NormalTok{rg\_mi }\OtherTok{\textless{}{-}} \FunctionTok{ref\_grid}\NormalTok{(fit\_mi, }
  \AttributeTok{at =} \FunctionTok{list}\NormalTok{(}
    \AttributeTok{Month =} \DecValTok{54}\NormalTok{,}
    \AttributeTok{active =} \FunctionTok{c}\NormalTok{(}\DecValTok{0}\NormalTok{,}\DecValTok{1}\NormalTok{)),}
  \AttributeTok{mode =} \StringTok{"satterthwaite"}\NormalTok{)}

\NormalTok{contrast\_mi }\OtherTok{\textless{}{-}} \FunctionTok{emmeans}\NormalTok{(rg\_mi, }\AttributeTok{specs =} \StringTok{\textquotesingle{}active\textquotesingle{}}\NormalTok{, }\AttributeTok{by =} \StringTok{\textquotesingle{}Month\textquotesingle{}}\NormalTok{, }\AttributeTok{lmerTest.limit=}\DecValTok{2000}\NormalTok{) }\SpecialCharTok{\%\textgreater{}\%}
  \FunctionTok{pairs}\NormalTok{(}\AttributeTok{reverse=}\ConstantTok{TRUE}\NormalTok{)}
\end{Highlighting}
\end{Shaded}

\hypertarget{treatment-effect-estimated-by-ncs-with-observed-data-only}{%
\subsection{Treatment effect estimated by NCS with observed data
only}\label{treatment-effect-estimated-by-ncs-with-observed-data-only}}

\begin{Shaded}
\begin{Highlighting}[]
\NormalTok{contrast\_observed}
\NormalTok{Month }\OtherTok{=} \DecValTok{54}\SpecialCharTok{:}
\NormalTok{ contrast          estimate    SE   df t.ratio p.value}
\NormalTok{ active1 }\SpecialCharTok{{-}}\NormalTok{ active0     }\FloatTok{1.16} \FloatTok{0.172} \DecValTok{7574}   \FloatTok{6.753}  \SpecialCharTok{\textless{}}\NormalTok{.}\DecValTok{0001}

\NormalTok{Results are averaged over the levels of}\SpecialCharTok{:}\NormalTok{ APOE4, version }
\NormalTok{Degrees}\SpecialCharTok{{-}}\NormalTok{of}\SpecialCharTok{{-}}\NormalTok{freedom method}\SpecialCharTok{:}\NormalTok{ satterthwaite }
\end{Highlighting}
\end{Shaded}

\hypertarget{treatment-effect-estimated-by-ncs-with-multiple-imputations}{%
\subsection{Treatment effect estimated by NCS with multiple
imputations}\label{treatment-effect-estimated-by-ncs-with-multiple-imputations}}

\begin{Shaded}
\begin{Highlighting}[]
\NormalTok{contrast\_mi}
\NormalTok{Month }\OtherTok{=} \DecValTok{54}\SpecialCharTok{:}
\NormalTok{ contrast          estimate    SE   df t.ratio p.value}
\NormalTok{ active1 }\SpecialCharTok{{-}}\NormalTok{ active0     }\FloatTok{1.02} \FloatTok{0.185} \DecValTok{7725}   \FloatTok{5.496}  \SpecialCharTok{\textless{}}\NormalTok{.}\DecValTok{0001}

\NormalTok{Results are averaged over the levels of}\SpecialCharTok{:}\NormalTok{ APOE4, version }
\NormalTok{Degrees}\SpecialCharTok{{-}}\NormalTok{of}\SpecialCharTok{{-}}\NormalTok{freedom method}\SpecialCharTok{:}\NormalTok{ satterthwaite }
\end{Highlighting}
\end{Shaded}

\hypertarget{tipping-point-analysis}{%
\section{Tipping point analysis}\label{tipping-point-analysis}}

\begin{Shaded}
\begin{Highlighting}[]
\NormalTok{post }\OtherTok{\textless{}{-}}\NormalTok{ result}\SpecialCharTok{$}\StringTok{\textasciigrave{}}\AttributeTok{2l.pmm}\StringTok{\textasciigrave{}}\SpecialCharTok{$}\NormalTok{post}
\NormalTok{k\_tipping }\OtherTok{\textless{}{-}} \FunctionTok{seq}\NormalTok{(}\FloatTok{4.25}\NormalTok{, }\FloatTok{5.25}\NormalTok{, }\FloatTok{0.25}\NormalTok{)}
\NormalTok{est\_tipping }\OtherTok{\textless{}{-}} \FunctionTok{vector}\NormalTok{(}\StringTok{"list"}\NormalTok{, }\FunctionTok{length}\NormalTok{(k\_tipping))}
\NormalTok{d }\OtherTok{\textless{}{-}}\NormalTok{ pad\_covid\_augment[, }\FunctionTok{c}\NormalTok{(}\StringTok{"id"}\NormalTok{, }\StringTok{"Month"}\NormalTok{, }\StringTok{"active"}\NormalTok{, }\StringTok{"APOE4"}\NormalTok{, }\StringTok{"age"}\NormalTok{, }\StringTok{"version"}\NormalTok{, }\StringTok{"Y"}\NormalTok{)]}
\NormalTok{pred }\OtherTok{\textless{}{-}} \FunctionTok{make.predictorMatrix}\NormalTok{(d)}
\ControlFlowTok{for}\NormalTok{ (k }\ControlFlowTok{in} \DecValTok{1}\SpecialCharTok{:}\FunctionTok{length}\NormalTok{(k\_tipping))\{}
  \CommentTok{\# decrease imputed PACC scores in the active group by }
  \CommentTok{\# k\_tipping[k] PACC points}
  \CommentTok{\# (nullify the imputed treatment effect to varying degrees)}
\NormalTok{  post[}\StringTok{"Y"}\NormalTok{] }\OtherTok{\textless{}{-}} \FunctionTok{paste}\NormalTok{(}\StringTok{"imp[[j]][,i] \textless{}{-} imp[[j]][,i] {-} "}\NormalTok{, k\_tipping[k], }
    \StringTok{" * data$active[!r[,j]]"}\NormalTok{)}
\NormalTok{  pred[, }\StringTok{"id"}\NormalTok{] }\OtherTok{\textless{}{-}} \SpecialCharTok{{-}}\DecValTok{2}
\NormalTok{  pred[}\StringTok{"id"}\NormalTok{, }\StringTok{"id"}\NormalTok{] }\OtherTok{\textless{}{-}} \DecValTok{0}
\NormalTok{  imp\_k }\OtherTok{\textless{}{-}} \FunctionTok{mice}\NormalTok{(d, }\AttributeTok{post=}\NormalTok{post, }\AttributeTok{predictorMatrix =}\NormalTok{ pred, }\AttributeTok{method =} \StringTok{"2l.pmm"}\NormalTok{,}
    \AttributeTok{m =} \DecValTok{10}\NormalTok{, }\AttributeTok{maxit =} \DecValTok{1}\NormalTok{, }\AttributeTok{print =} \ConstantTok{FALSE}\NormalTok{, }\AttributeTok{seed =} \DecValTok{82828}\NormalTok{)}
\NormalTok{  fit\_k }\OtherTok{\textless{}{-}} \FunctionTok{with}\NormalTok{(imp\_k, }\FunctionTok{lmer}\NormalTok{(Y }\SpecialCharTok{\textasciitilde{}} 
      \FunctionTok{I}\NormalTok{(}\FunctionTok{ns21}\NormalTok{(Month)) }\SpecialCharTok{+} \FunctionTok{I}\NormalTok{(}\FunctionTok{ns22}\NormalTok{(Month)) }\SpecialCharTok{+}
\NormalTok{      (}\FunctionTok{I}\NormalTok{(}\FunctionTok{ns21}\NormalTok{(Month)) }\SpecialCharTok{+} \FunctionTok{I}\NormalTok{(}\FunctionTok{ns22}\NormalTok{(Month)))}\SpecialCharTok{:}\NormalTok{active }\SpecialCharTok{+} 
\NormalTok{      APOE4 }\SpecialCharTok{+}\NormalTok{ age }\SpecialCharTok{+}\NormalTok{ version }\SpecialCharTok{+}\NormalTok{ (}\DecValTok{1} \SpecialCharTok{|}\NormalTok{ id)))}
\NormalTok{  rg\_k }\OtherTok{\textless{}{-}} \FunctionTok{ref\_grid}\NormalTok{(fit\_k, }
    \AttributeTok{at =} \FunctionTok{list}\NormalTok{(}
      \AttributeTok{Month =} \DecValTok{54}\NormalTok{,}
      \AttributeTok{active =} \FunctionTok{c}\NormalTok{(}\DecValTok{0}\NormalTok{,}\DecValTok{1}\NormalTok{)),}
    \AttributeTok{mode =} \StringTok{"satterthwaite"}\NormalTok{)}
\NormalTok{  ct\_k }\OtherTok{\textless{}{-}} \FunctionTok{emmeans}\NormalTok{(rg\_k, }\AttributeTok{specs =} \StringTok{\textquotesingle{}active\textquotesingle{}}\NormalTok{, }\AttributeTok{by =} \StringTok{\textquotesingle{}Month\textquotesingle{}}\NormalTok{, }\AttributeTok{lmerTest.limit=}\DecValTok{2000}\NormalTok{) }\SpecialCharTok{\%\textgreater{}\%}
    \FunctionTok{pairs}\NormalTok{(}\AttributeTok{reverse=}\ConstantTok{TRUE}\NormalTok{) }\SpecialCharTok{\%\textgreater{}\%}
    \FunctionTok{as.data.frame}\NormalTok{()}
\NormalTok{  est\_tipping[[k]] }\OtherTok{\textless{}{-}}\NormalTok{ ct\_k}
\NormalTok{\}}

\NormalTok{results\_tipping }\OtherTok{\textless{}{-}} \FunctionTok{bind\_rows}\NormalTok{(est\_tipping)}
\NormalTok{results\_tipping }\OtherTok{\textless{}{-}} \FunctionTok{bind\_cols}\NormalTok{(}\AttributeTok{tipping\_factor =}\NormalTok{ k\_tipping, results\_tipping)}
\NormalTok{results\_tipping }\SpecialCharTok{\%\textgreater{}\%}
\NormalTok{  knitr}\SpecialCharTok{::}\FunctionTok{kable}\NormalTok{(}\AttributeTok{digits=}\DecValTok{3}\NormalTok{, }\AttributeTok{caption =} \StringTok{\textquotesingle{}Treatment effects for different MNAR factors.\textquotesingle{}}\NormalTok{)}
\end{Highlighting}
\end{Shaded}

\begin{longtable}[]{@{}
  >{\raggedleft\arraybackslash}p{(\columnwidth - 14\tabcolsep) * \real{0.1899}}
  >{\raggedright\arraybackslash}p{(\columnwidth - 14\tabcolsep) * \real{0.2278}}
  >{\raggedleft\arraybackslash}p{(\columnwidth - 14\tabcolsep) * \real{0.0759}}
  >{\raggedleft\arraybackslash}p{(\columnwidth - 14\tabcolsep) * \real{0.1139}}
  >{\raggedleft\arraybackslash}p{(\columnwidth - 14\tabcolsep) * \real{0.0759}}
  >{\raggedleft\arraybackslash}p{(\columnwidth - 14\tabcolsep) * \real{0.1139}}
  >{\raggedleft\arraybackslash}p{(\columnwidth - 14\tabcolsep) * \real{0.1013}}
  >{\raggedleft\arraybackslash}p{(\columnwidth - 14\tabcolsep) * \real{0.1013}}@{}}
\caption{Treatment effects for different MNAR factors.}\tabularnewline
\toprule()
\begin{minipage}[b]{\linewidth}\raggedleft
tipping\_factor
\end{minipage} & \begin{minipage}[b]{\linewidth}\raggedright
contrast
\end{minipage} & \begin{minipage}[b]{\linewidth}\raggedleft
Month
\end{minipage} & \begin{minipage}[b]{\linewidth}\raggedleft
estimate
\end{minipage} & \begin{minipage}[b]{\linewidth}\raggedleft
SE
\end{minipage} & \begin{minipage}[b]{\linewidth}\raggedleft
df
\end{minipage} & \begin{minipage}[b]{\linewidth}\raggedleft
t.ratio
\end{minipage} & \begin{minipage}[b]{\linewidth}\raggedleft
p.value
\end{minipage} \\
\midrule()
\endfirsthead
\toprule()
\begin{minipage}[b]{\linewidth}\raggedleft
tipping\_factor
\end{minipage} & \begin{minipage}[b]{\linewidth}\raggedright
contrast
\end{minipage} & \begin{minipage}[b]{\linewidth}\raggedleft
Month
\end{minipage} & \begin{minipage}[b]{\linewidth}\raggedleft
estimate
\end{minipage} & \begin{minipage}[b]{\linewidth}\raggedleft
SE
\end{minipage} & \begin{minipage}[b]{\linewidth}\raggedleft
df
\end{minipage} & \begin{minipage}[b]{\linewidth}\raggedleft
t.ratio
\end{minipage} & \begin{minipage}[b]{\linewidth}\raggedleft
p.value
\end{minipage} \\
\midrule()
\endhead
4.25 & active1 - active0 & 54 & 0.466 & 0.187 & 7680.640 & 2.492 &
0.013 \\
4.50 & active1 - active0 & 54 & 0.433 & 0.187 & 7676.295 & 2.316 &
0.021 \\
4.75 & active1 - active0 & 54 & 0.401 & 0.187 & 7671.759 & 2.140 &
0.032 \\
5.00 & active1 - active0 & 54 & 0.368 & 0.187 & 7667.035 & 1.965 &
0.049 \\
5.25 & active1 - active0 & 54 & 0.336 & 0.188 & 7662.122 & 1.790 &
0.073 \\
\bottomrule()
\end{longtable}

The NCS model estimate of a 1.16 PACC point effect is robust to
perturbations of the MAR imputations in the active group toward the null
up to about 5 PACC points, or 5/1.16 = 4.31 times the NCS estimated
effect.

\hypertarget{references}{%
\section*{References}\label{references}}
\addcontentsline{toc}{section}{References}

\hypertarget{refs}{}
\begin{CSLReferences}{1}{0}
\leavevmode\vadjust pre{\hypertarget{ref-R-lme4}{}}%
Bates, Douglas, Martin Maechler, Ben Bolker, and Steven Walker. 2022.
\emph{{lme4}: Linear Mixed-Effects Models Using Eigen and S4}.
\url{https://github.com/lme4/lme4/}.

\leavevmode\vadjust pre{\hypertarget{ref-donohue2014preclinical}{}}%
Donohue, Michael C, Reisa A Sperling, David P Salmon, Dorene M Rentz,
Rema Raman, Ronald G Thomas, Michael Weiner, Paul S Aisen, et al. 2014.
{``The Preclinical Alzheimer Cognitive Composite: Measuring
Amyloid-Related Decline.''} \emph{JAMA Neurology} 71 (8): 961--70.

\leavevmode\vadjust pre{\hypertarget{ref-R-pan}{}}%
Joseph L. Schafer, Original by. 2018. \emph{{pan}: Multiple Imputation
for Multivariate Panel or Clustered Data}.
\url{https://CRAN.R-project.org/package=pan}.

\leavevmode\vadjust pre{\hypertarget{ref-R-emmeans}{}}%
Lenth, Russell V. 2022. \emph{{emmeans}: Estimated Marginal Means, Aka
Least- Squares Means}. \url{https://github.com/rvlenth/emmeans}.

\leavevmode\vadjust pre{\hypertarget{ref-R-miceadds}{}}%
Robitzsch, Alexander, Simon Grund, and Thorsten Henke. 2022.
\emph{{miceadds}: Some Additional Multiple Imputation Functions,
Especially for Mice}. \url{https://CRAN.R-project.org/package=miceadds}.

\leavevmode\vadjust pre{\hypertarget{ref-rubin1977formalizing}{}}%
Rubin, Donald B. 1977. {``Formalizing Subjective Notions about the
Effect of Nonrespondents in Sample Surveys.''} \emph{Journal of the
American Statistical Association} 72 (359): 538--43.

\leavevmode\vadjust pre{\hypertarget{ref-van2018flexible}{}}%
van Buuren, Stef. 2018. \emph{Flexible Imputation of Missing Data}. CRC
press. \url{https://stefvanbuuren.name/fimd/}.

\leavevmode\vadjust pre{\hypertarget{ref-mice2011}{}}%
van Buuren, Stef, and Karin Groothuis-Oudshoorn. 2011. {``{mice}:
Multivariate Imputation by Chained Equations in r.''} \emph{Journal of
Statistical Software} 45 (3): 1--67.
\url{https://doi.org/10.18637/jss.v045.i03}.

\leavevmode\vadjust pre{\hypertarget{ref-R-mice}{}}%
---------. 2021. \emph{{mice}: Multivariate Imputation by Chained
Equations}. \url{https://CRAN.R-project.org/package=mice}.

\leavevmode\vadjust pre{\hypertarget{ref-tidyverse2019}{}}%
Wickham, Hadley, Mara Averick, Jennifer Bryan, Winston Chang, Lucy
D'Agostino McGowan, Romain François, Garrett Grolemund, et al. 2019.
{``Welcome to the {tidyverse}.''} \emph{Journal of Open Source Software}
4 (43): 1686. \url{https://doi.org/10.21105/joss.01686}.

\end{CSLReferences}